\documentclass[conference]{IEEEtran}
\IEEEoverridecommandlockouts
\usepackage{cite}
\usepackage{amsmath,amssymb,amsfonts}
\usepackage{algorithmic}
\usepackage{graphicx}
\usepackage{textcomp}
\usepackage{xcolor}

\usepackage{amsmath,amssymb,amsfonts}
\usepackage{algorithmic}
\usepackage{graphicx}
\usepackage{textcomp}
\usepackage{color,soul}
 \usepackage{multirow}
 \usepackage{subcaption}
 \usepackage{enumitem}
 \usepackage{mathtools}

\usepackage{braket}

\usepackage{caption}

\usepackage{url}

\usepackage{hyperref}

\def\BibTeX{{\rm B\kern-.05em{\sc i\kern-.025em b}\kern-.08em
    T\kern-.1667em\lower.7ex\hbox{E}\kern-.125emX}}
\begin{document}

\title{QuFI: a Quantum Fault Injector to Measure the Reliability of Qubits and Quantum Circuits}


\author{
\IEEEauthorblockN{Daniel Oliveira\IEEEauthorrefmark{2}, Edoardo Giusto\IEEEauthorrefmark{1},
Emanuele Dri\IEEEauthorrefmark{1},
Nadir Casciola\IEEEauthorrefmark{1},\\
Betis Baheri\IEEEauthorrefmark{3}, Qiang Guan\IEEEauthorrefmark{3}, 
Bartolomeo Montrucchio\IEEEauthorrefmark{1}, Paolo Rech\IEEEauthorrefmark{4}}\\

\IEEEauthorblockA{\IEEEauthorrefmark{2}Department of Informatics, Federal University of Paran\'{a} (UFPR), Curitiba, Brazil\\
dagoliveira@inf.ufpr.br
}
\IEEEauthorblockA{\IEEEauthorrefmark{1}DAUIN, Politecnico di Torino, Torino, Italy\\
\{edoardo.giusto, bartolomeo.montrucchio\}@polito.it\\
\{emanuele.dri, nadir.casciola\}@studenti.polito.it
}
\IEEEauthorblockA{\IEEEauthorrefmark{3}Department of Computer Science, Kent State University, Kent, USA\\
\{bbaheri, qguan\}@kent.edu 
}
\IEEEauthorblockA{\IEEEauthorrefmark{4}Department of Industrial Engineering, Universit\`a di Trento, Italy\\
paolo.rech@unitn.it
}
}
\maketitle

\thispagestyle{plain}
\pagestyle{plain}

\begin{abstract}
Quantum computing is an up-and-coming technology that is expected to revolutionize the computation paradigm in the next few years.
Qubits, the primary computing elements of quantum circuits, exploit the quantum physics proprieties to increase the parallelism and speed of computation drastically.
Unfortunately, besides being intrinsically noisy, qubits have also been shown to be highly susceptible to external sources of faults, such as ionizing radiation.
The latest discoveries highlight a much higher radiation sensitivity of qubits than traditional transistors and identify a much more complex fault model than bit-flip.

We propose a framework to identify the quantum circuits sensitivity to radiation-induced faults and the probability for a fault in a qubit to propagate to the output. 
Based on the latest studies and radiation experiments performed on real quantum machines, we model the transient faults in a qubit as a phase shift with a parametrized magnitude.
Additionally, our framework can inject multiple qubit faults, tuning the phase shift magnitude based on the proximity of the qubit to the particle strike location.
As we show in the paper, the proposed fault injector is highly flexible, and it can be used on both quantum circuit simulators and real quantum machines.
We report the finding of more than $285,249,536$ injections on the Qiskit simulator and $53,248$ injections on real IBM machines. We consider three quantum algorithms and identify the faults and qubits that are more likely to impact the output.
We also consider the fault propagation dependence on the circuit scale, showing that the reliability profile for some quantum algorithms is scale-dependent, with increased impact from radiation-induced faults as we increase the number of qubits. Finally, we also consider multi qubits faults, showing that they are much more critical than single faults.
The fault injector and the data presented in this paper are available in a public repository to allow further analysis.
\end{abstract}



\begin{IEEEkeywords}
component, formatting, style, styling, insert
\end{IEEEkeywords}

\section{Introduction}
\label{sec_intro}

Being one of the most promising technology advances to move beyond the CMOS transistor technology and Moore's law, Quantum Computing (QC) has attracted significant interest from the research community. In the last few years, QC has moved from being an attractive conceptual solution for physics problems to a highly parallel and highly efficient computing architecture for various applications~\cite{1996Grover, lloyd2013quantum, Peruzzo2014, quantum-drug}. The availability of the first quantum computers and the development of easy to use simulators and frameworks have suddenly raised the community's interest in quantum computing. Moreover, the billions of dollars investments of industries, research centers, and government agencies are encouraging the development of large-scale quantum computers and the training of quantum programmers and designers.

Qubits, the fundamental computing element of a quantum circuit, have an intrinsic reliability vulnerability that comes from their sensitivity to noise and external perturbations. The intriguing theory of quantum computing became a promising computing paradigm when a sufficiently stable and fault-tolerant qubit to allow the computation of small yet crucial circuits was finally produced~\cite{Bravyi2018, Chamberland_2020}.
Today, researchers have access to several prototypes of real quantum machines, such as IBM, D-Wave, Rigetti, Pasqal, and quantum circuit simulators~\cite{Qiskit, AngliQSimulator}.

Recent inspiring publications have shown that superconducting qubits are particularly susceptible also to external radiation~\cite{radiation2011, Cardani2021, Martinis2021, Chen2021}. Unfortunately, it was shown that the impact of ionizing particles undermines the achieved stability and noise tolerance of qubits~\cite{LossMechanisms2018, nature_rad, muons2021}. 
Ionizing radiation, which is one of the biggest challenges for modern classical computing systems, is expected to be a major issue also for future quantum (super) computers~\cite{Cardani2021, Chen2021}.
The reliability of quantum computers to radiation seems exceptionally challenging as the latest discoveries highlighted that qubits have an even higher sensitivity to external perturbation than CMOS transistors and that qubits state is also modified by light particles, such as muons~\cite{muons2021} or even infrared light~\cite{Barends2011}, that are harmless to CMOS devices.

Besides having a possibly higher sensitivity to radiation, qubits also have a much more complex fault model than CMOS devices.
The transistor has a binary state (ON or OFF), and a fault is generated when the deposited charge is sufficient to trigger a transistor state change.
The fault in a CMOS device is then binary.
Qubits state is not binary and is represented on the Block sphere (see Figure~\ref{fig:BlochSphere}).
A small deposited charge can modify the qubit state and eventually propagate to the circuit output. 
A challenging aspect of qubits reliability is that any perturbation can change the state and the state change is not binary. This latter aspect is particularly challenging when designing a fault injector for the quantum circuit, like the one we propose in this paper. The fault model to inject is not simply a transition from ON to OFF or from 1 to 0, but rather a phase shift of different directions and magnitudes.

This paper formalizes the transient fault model for qubits based on recent discoveries on quantum materials' radiation effects. Then, we present the first fault injector designed to estimate the reliability of transient faults of a quantum circuit. To analyze qubits and quantum circuit reliability, we use a new metric based on Michelson Contrast, the \textit{Quantum Vulnerability Factor} (QVF), clearly inspired by the Architectural or Program Vulnerability Factors (AVF or PVF)~\cite{Mukherjee2003, PVF}. We perform single and multi-qubit fault injection, identifying the criticality of different faults and the more sensitive qubits to radiation-induced faults. Such information is precious as it allows a reliability-aware mapping of the circuit qubits to physical qubits, predicts the effects of faults in the quantum computation, and focuses the eventual additional fault tolerance solution to the most critical qubit(s). We also evaluate the reliability impact when scaling a circuit (i.e., increasing the number of qubits and operations performed). We have found that some circuits present different reliability profiles that are scale-dependent, and if not mitigated, will hinder the applicability of quantum circuits to real problems that require large scale. Finally, we compare the results obtained on the Qiskit simulator with fault injection performed on real IBM machines, showing that the simulator evaluation is accurate.

The rest of the paper is organized as follows.
To have a basic notion about quantum computers and prove the importance of understanding the reliability of quantum circuits, we give, in Section~\ref{sec_background}, some background information about quantum computing and radiation effects in qubits.
Then, in Section~\ref{sec_model}, we describe how we model transient faults using the results of recently performed experiments on qubits materials. In Section~\ref{sec_fi}, we present our fault injector and formalize the metric we use to understand the impact of faults in quantum circuits.
The obtained results are presented in Section~\ref{sec_results},  and Section~\ref{sec_conclusion} concludes the paper.





\section{Background and Related Work}
\label{sec_background}


In this Section, we give a background about qubits, quantum computing, and the quantum programming, which is essential to understand and appreciate the contribution of our paper. We will also describe the current reliability challenges of qubits, the available fault tolerance solutions, and their limitations.

There are various technologies available to implement qubits, the most promising ones being superconducting qubits and trapped-ions qubits. As there is not yet sufficient data on radiation effects on trapped-ion qubits (besides the low-dose test documented in~\cite{trappedionlowdose}), we will focus the discussion on superconducting materials. Nonetheless, the concepts we introduce, the fault injection framework we design, and the impact of the results we present are independent on the qubit technology, once the fault model is defined.

\subsection{Basic notions of quantum computing}
\label{sub:basic}
\begin{figure}[t!]
    \includegraphics[width=0.4\textwidth]{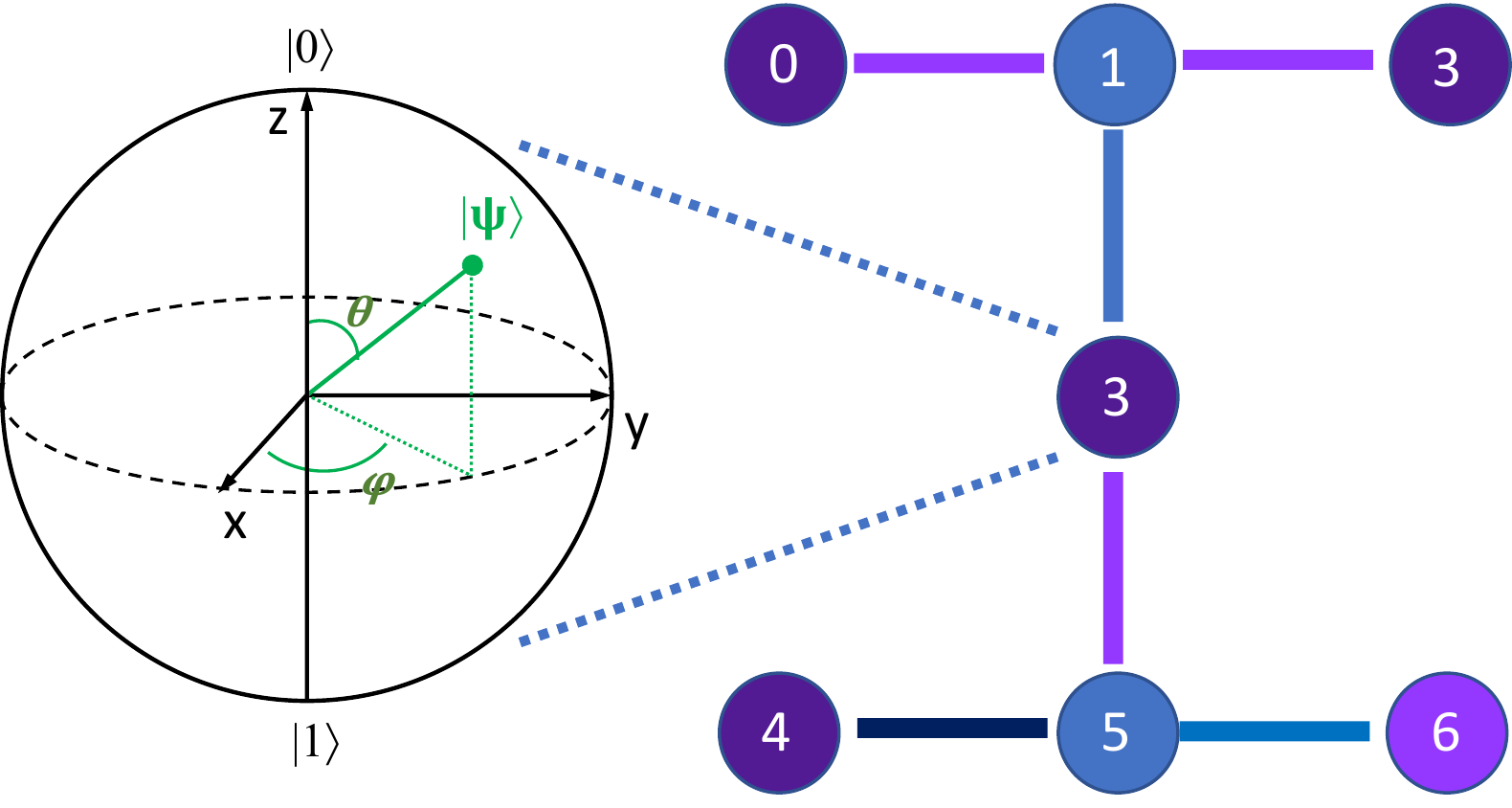}
    \centering
    \caption{Bloch sphere that visualizes the state of a qubit (left) and how qubits are connected in the Casablanca architecture of IBM (right).}
    \label{fig:BlochSphere}
\end{figure}

Traditional, classical, computers perform operations through the use of bits.
The state of each bit can be either a deterministic $0$ or $1$. 
In quantum computers, instead, the computation is done on quantum bits (qubits).
A qubit in general is not in one \textit{classical} state but in a \textit{quantum} state.
A qubit can exist in a \textit{superposition} of quantum states, denoted as:
\begin{math} \ket{\Psi} = \alpha \ket{0} + \beta \ket{1}
\end{math}
, where $\ket{\Psi}$ is the actual qubit state, $\alpha$ and $\beta$ are complex numbers that represent the probability amplitude of $\ket{0}$ or $\ket{1}$ state.
Visually, the qubit state $\ket{\Psi}$ is represented as a vector on the Block Sphere, shown on the left of Figure~\ref{fig:BlochSphere}, and is described by
the two angles $\theta$ and $\phi$.
The state of a qubit, and then the quantum computation output, is \textit{probabilistic} and not deterministic.

Quantum gates operate on qubits and can put a qubit in any superposition of $\ket{0}$ and $\ket{1}$ (change the probability of 0 and 1). 
There are many available quantum gates, such as Hadamard, Pauli-X, Pauli-Y, Pauli-Z, Phase, etc.
The Hadamard (H) gate puts a qubit in an equal superposition of $\ket{0}$ and $\ket{1}$, the X gate (or the bit-flip gate) performs a rotation of $90^\circ$ about the x-axis on the Bloch Sphere, and the Z gate (or the phase-flip gate) performs a rotation of $90^\circ$ about the z-axis. 

Quantum programs are expressed as quantum circuits, which are a set of quantum gates that are sequentially applied to the initial qubits, and produce a probabilistic output(s).
These quantum circuits are mapped on Noisy Intermediate Scale Quantum Computers (NISQ) machines based on qubits connectivity and supported gates. NISQ machines can have different topologies, with some qubits that can be directly accessible.  For example, as shown in Figure~\ref{fig:BlochSphere}, on Casablanca qubit 0 and 1 are directly connected and can interact with each other.
Then, multi qubit gates can be applied to qubits 0 and 1, which is not the case for qubits without a direct connection. 

To be executed, a quantum circuit needs to be mapped to the target architecture.
\textit{Transpiling} is the process of implementing the quantum circuit, i.e., assigning the logical qubits to the available physical locations and adding the quantum gates that execute the operations.
Transpiling corresponds to the compiling process for source codes or the synthesis for Hardware Description Language in classical computation. 
A higher number of qubits allows to work on more complex and realistic workloads, to improve performances, and guarantee higher success probability rate. Industry is working to increase the number of stable qubits, with IBM already having a 127 qubits machine. As part of our evaluation, we investigate if and how the scale of the circuit (i.e., the number of qubits and gates applied) influences its reliability to transient faults.

Currently, IBM, the most widely used provider, disposes of 23 quantum computers. 
Each quantum processor model uses different qubit topologies, which may greatly impact the machine quantum noise~\cite{qucloudHPCA}. 
Additionally, frameworks have been developed for the simulation and the use of real machines, to ease the research and the training of quantum programmers.
Qiskit~\cite{Qiskit} is designed to be machine-independent and it is the first released and the most used one. 
In our experiments, we mainly focus on studying the IBM quantum machines and running benchmark quantum programs in Qiskit.

\subsection{Quantum noise and fault tolerance}
\label{sec_noise}

The performance and reliability of quantum computers are bounded by the intrinsic noise, which greatly reduces the accuracy of quantum computation~\cite{Harper2020}.
Quantum noise can be categorized into operation errors and retention (coherence) errors~\cite{2017APSMARR51007G}. A qubit can properly maintain its state (data) for a limited time (coherence time). The qubit state degradation is called retention error and is categorized into two types, T1 and T2 errors~\cite{Hu01Decoherence}.

A qubit in a high energy state $\ket{1}$ naturally decays to lower energy state $\ket{0}$, the time associated with this intrinsic decay is called spin-lattice coherence time (T1). The spin-spin relaxation process, on the contrary, indicates the time (T2) for external environment or for the interaction with other qubits to affect the qubit state. 
Depending on the technology, individual qubits have a time range for T1 and T2 that has improved, in the last decade, from 1 nano-second to 100 micro-seconds~\cite{Kjaergaard_2020}.

To reduce the quantum noise, industries and researchers are working to improve qubit and the machine design, even using new isolation techniques and separating the quantum computer from the surrounding environment.
The hardware approach is only one of the dimensions to consider. Quantum Error Correction (QEC) is adopted to restructure the quantum circuits, adding redundancy to improve the resilience to noise and ensure a sufficiently long computation time. QEC has been fundamental to reach Fault Tolerant Quantum Computers (FTQC) but is extremely costly, as it requires from 5x to 9x larger circuits~\cite{qce-2021}. 

It is worth noting that QEC is designed to protect a qubit from the intrinsic noise, which is well studied and predictable.
Unfortunately, as we discuss next, QEC is inefficient in handling radiation-induced transient faults~\cite{nature_rad, muons2021, LossMechanisms2018}. To reach FTQC we need to design better QEC, and the first step is the understanding of faults impact and propagation, which is the main scope of this paper.


\subsection{Radiation-Induced Faults in qubits}
\label{sub_radiation}

\begin{figure*}[t]%
   	\centering
    	\includegraphics[width=0.88\textwidth]{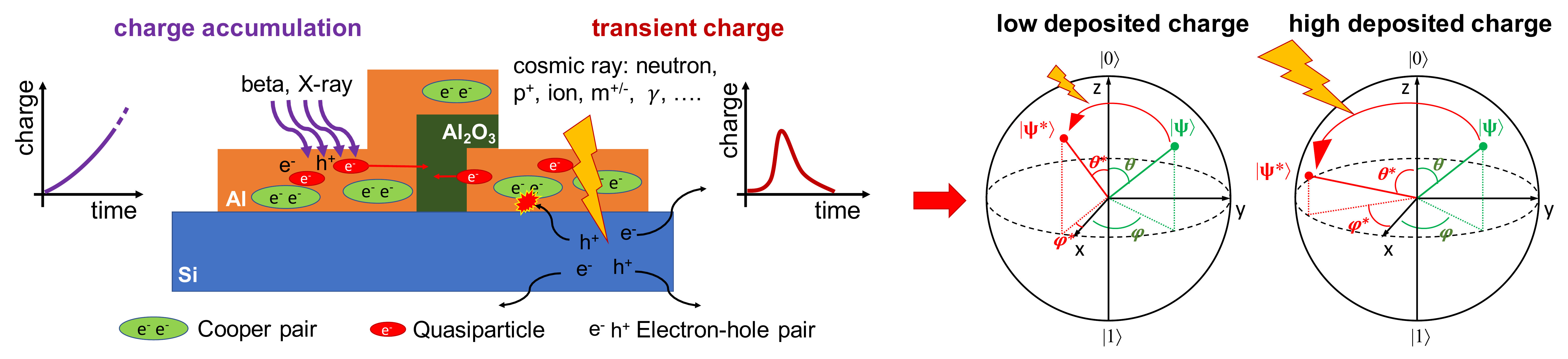}
        \caption{Simplified effect of the impact of accumulated (from beta and x-rays) and transient (from cosmic rays) charge deposition in a superconducting qubit material (on the left, adapted from~\cite{nature_rad}) and its consequences on the qubit state (on the right, adapted from~\cite{muons2021}). The charge deposited by radiation in Silicon or Aluminum generates electron-hole pairs that induce a non-equilibrium and breaks the Cooper pair, producing quasiparticles. The excess of quasiparticles excites the qubit. The excitement is logically translated into phase(s) shift(s) with a magnitude that depends on the deposited charge and a consequent state change from the expected $| \Psi \rangle$ to the corrupted $| \Psi ^\ast \rangle$.}
        \label{fig_rad_effect}
\end{figure*}

Preliminary and inspiring works show that ionizing radiation induces faults in superconducting qubits~\cite{LossMechanisms2018, nature_rad, muons2021, Barends2011} and, once employed in large scale, radiation fault tolerance is expected to be the next big challenge for quantum (super-) computers~\cite{radiation2011, Cardani2021, Martinis2021, Chen2021}. 
Lately it has been shown that the frailty of qubits makes it possible also for low energy neutrons and lighter particles, such as muons (almost harmless for CMOS technology~\cite{muons2011}), to induce a state perturbation~\cite{muons2021}. Intuitively, the charge deposited by the impinging particle excites the qubit, modifying its state and, thus, generating a fault, as shown in Figure~\ref{fig_rad_effect} and detailed in Section~\ref{sec_model}. This also implies that shielding, already impractical for CMOS devices, becomes impossible. As shown in~\cite{muons2011}, only putting qubits in caves under mountains can eventually reduce the muons effects on the quantum state.

Unfortunately, Quantum Error Correction, that has enabled the scaling of quantum computers, is designed to be effective for the noise, not for transient faults.
Noise is a well studied, well characterized and predictable phenomenon in quantum circuits. IBM, for instance, releases a daily report about the noise level of their machines. 
Transient faults are stochastic, unpredictable, and add over the intrinsic noise. 
As shown in~\cite{radiation2011, Cardani2021, Martinis2021, Chen2021}, current QEC is not sufficient to guarantee reliability from transient faults.
Given the high cost of QEC, it is fundamental to understand if and how a transient fault propagates in the quantum circuits to design effective and efficient hardening solutions, as is currently being done for traditional electronic systems.
A fault injection for transient faults in qubits is a powerful tool to move in this direction.



\subsection{Contribution}
\label{sec_contrib}

In this paper, we address three main challenges associated with the reliability evaluation of quantum circuits: (1) the formalization of qubit(s) fault model, (2) the design of a flexible framework to inject errors, and (3) a systematic methodology to quantify and qualify the impact of transient fault in the quantum circuit output. 

We first formalize the transient fault model for qubits, based on the most recent discoveries on radiation effects in quantum materials. Then, we present QuFI, the fault injector aimed at practically estimating the reliability of a quantum circuit and at identifying the qubits that, if affected, are more likely to induce a negative impact on the output correctness. Such information is highly valuable as it allows a reliability-aware mapping of the circuit qubits to physical qubits, to predict the effects of faults in the quantum computation, and to focus the eventual additional fault tolerance solution to the most critical qubits. To analyze qubits and quantum circuit reliability, we use a new metric based on Michelson Contrast, the \textit{Quantum Vulnerability Factor} (QVF), clearly inspired by the Architectural or Program Vulnerability Factors (AVF or PVF)~\cite{Mukherjee2003, PVF}.

While previous work on quantum fault tolerance studied the noise and fault generation, we focus on the effects of faults propagation in quantum circuits. In other words, as in the AVF or PVF measurement for traditional computation, we assume that the fault occurred and study its effect on the quantum circuit output.
Recently a preliminary fault injector to track noise propagation was presented~\cite{Resch2021}. However, to the best of our knowledge, ours is the first fault injector expressively designed to track transient fault propagation in quantum circuits. Additionally, our tool, unlike the available one, is integrated with Qiskit, uses a realistic noise and fault model, and, as we show, can also be used in real quantum machines. 

To have a broad evaluation of transient faults propagation in quantum circuits, we consider 3 widely-known circuits.
To understand the QVF dependence on the number of qubits we implement the circuits with increased complexity.
We also compare the reliability of circuits when single or multi qubit faults.





\section{Radiation-Induced Quantum Fault Model}
\label{sec_model}

In this section we describe how we model the radiation-induced faults in qubits. We use the available knowledge on particles interaction with qubits materials and the results of recently performed experiments and simulations to model the radiation-induced transient faults in quantum circuits. The details provided in this Section serve as a solid background to justify the chosen fault model.

\subsection{Single qubit fault}
\label{sub_single}

Superconducting qubits are built with Aluminum thin-film and a Silicon substrate, as shown in the left part of Figure~\ref{fig_rad_effect}, adapted from~\cite{nature_rad}.
It is well known, from the studies on semiconductor reliability~\cite{Baumann2005}, that the interaction of energetic particles with Silicon and other materials generates electron-hole (e-h) pairs.
In particular, heavy particles, such as the secondary production of galactic cosmic rays with the atmosphere, are more likely to interact with Silicon ($Al$ is transparent to neutrons), other sources of (light) radiation ($\beta$, X-rays) with Aluminum.

The non-equilibrium that results from the charge deposited by radiation in the qubit material induces Cooper pairs breaking and, thus, quasiparticles generation~\cite{nature_rad, muons2021}. The resulting excitement modifies the state of the qubit, possibly changing its state (i.e., it induces $\phi$ and/or $\theta$ phase shift) as shown with simulations~\cite{muons2021} and experimentally validated~\cite{Cardani2021}.


The major difference between CMOS technology and qubits, which represents one of the main challenges we address in the proposed fault model, is that, while a particle induces a fault in a CMOS transistor only if it deposits a charge higher than the critical one~\cite{Baumann2005}, on a qubit any excitement modifies the state, inducing a phase shift with a magnitude that depends on the deposited charge~\cite{Catelani2011}. In other words, in the binary CMOS technology, the particle either generates a fault or not. The "simple" bit-flip fault model is then sufficient to study CMOS devices reliability. 
On the contrary, on a qubit the magnitude of the fault (phase shift) depends on the amount of charge that reaches the qubit. It is then necessary to inject faults (phase shifts) of different magnitudes. 

Figure~\ref{fig_ion} shows a simplified example of the spatial distribution of the charge deposited in Silicon by a Fe ion of 275MeV (electronVolt). The simulation data was taken and adapted from GEANT4 simulations~\cite{Geant4} presented in~\cite{strike}, and serves as an illustrative example.
As shown, the amount of e-h pairs deposited exponentially decreases with distance. A qubit in the proximity of the particle impact will have a huge phase shift, while qubits further than $\sim1\mu m$ will be barely affected.
Thus, to study the sensitivity of a quantum circuit we need to inject various phase shifts of different magnitude. On a CMOS, only the transistor in the proximity of the particle strike would have been flipped.

It is worth noting that if, and only if, the deposited charge is sufficiently high the qubit can collapse~\cite{nature_rad}. 
However, as the phase shift magnitude depends on the deposited charge~\cite{Catelani2011} and the low energy neutrons are exponentially more common than high energy ones (the spectrum of neutrons ranges from meV to GeV~\cite{Baumann2005}), we can derive that collapses are less likely than phase shifts. In this work we do not consider qubit collapse as, when this happens, the quantum circuit ceases to exist and there is no fault propagation to consider.

\subsection{Transient and accumulative charge}
\label{sub_accumulation}

Similar to traditional CMOS, even for qubits two kind of radiation-induced effects have been highlighted: charge accumulation and transient charge deposition.

Terrestrial neutrons, ions, protons, and muons strike is stochastic and generates a \textit{transient} large amount of electrons-holes pairs in the Silicon substrate. As shown in Figure~\ref{fig_rad_effect} and detailed in~\cite{Baumann2005}, the deposited charge quickly increases for then slowly disappearing. This transient charge deposition is today one of the most challenging reliability issues for electronic devices. 
On the contrary, $\gamma$-rays, $\beta$, X-rays have a constant and \textit{accumulative} effect, known as Total Ionizing Dose (TID)~\cite{Oldham2003}. The exposure to $\gamma$-rays, $\beta$, and X-rays constantly deposits a little amount of charge that accumulates over time, as shown in Figure~\ref{fig_rad_effect}. When the deposited charge is too high it induces a permanent CMOS transistor malfunction~\cite{Oldham2003} or qubit collapse~\cite{nature_rad}.

Charge accumulation is only apparently more critical than transient charge deposition. In fact, TID effects are very rare in terrestrial applications as neutrons and muons, which are (by far) the most common particles, deposits negligible charge~\cite{Baumann2005}. TID are indeed a problem mainly for long lasting space applications~\cite{Oldham2003}, which are not (yet) interested in quantum computing. Additionally, a thin shielding is sufficient to drastically reduce the charge deposited by X-rays in a qubit, as shown in~\cite{nature_rad}. On the contrary, the shielding for neutrons and heavy ions is impractical (meters of concrete or lead) and for muons is basically impossible as the qubit should be placed in deep underground caves~\cite{muons2021}.
As already well established for traditional electronic devices, it is then also impossible to shield qubits from transient faults. We then focus our effort in studying the unavoidable transient faults propagation in quantum circuits, leaving the extension to accumulative charge effects as a future work.

\subsection{Multi-qubits faults}
\label{sub_multi}

\begin{figure}[t]%
   	\centering
    	\includegraphics[width=0.48\textwidth]{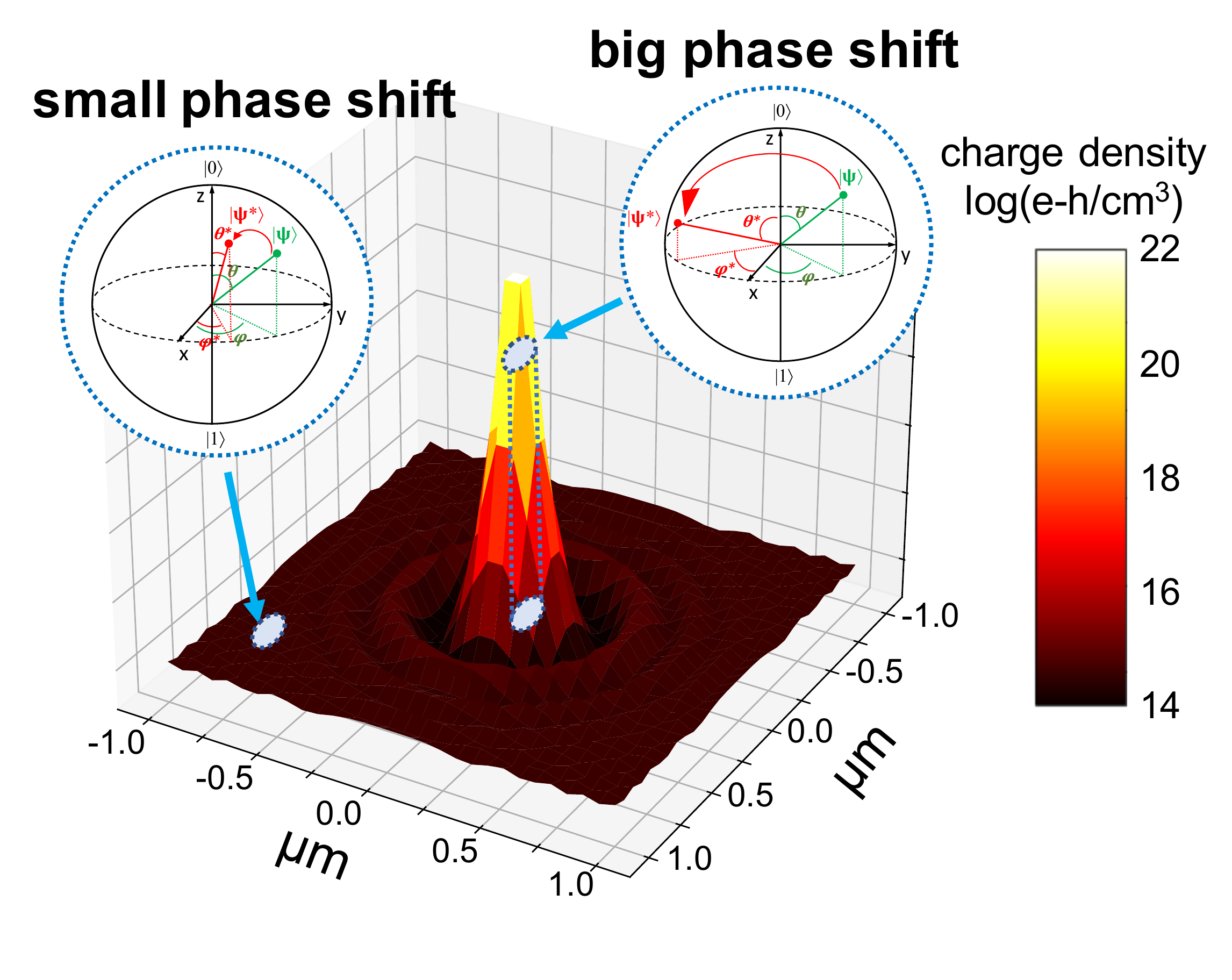}
        \caption{Simplified density of deposited charge (electron-hole pairs per $cm^3$) of a an impinging particle (275MeV) in Silicon, adapted from~\cite{strike}. The qubit closer to the particle strike will be affected by a phase shift of higher magnitude.}
        \label{fig_ion}
\end{figure}

As extensively shown in~\cite{muons2021}, the particle impact can interact with more than one qubit, generating multi-qubits faults. The relation between the phase shifts of the qubits that are affected by the same impinging particle is still largely unclear. Nonetheless, as depicted in Figure~\ref{fig_ion}, it is reasonable to assume that the affected qubits, being probably at different distances from the particle impact, will experience phase shifts of different magnitude. The qubit closer to the particle impact suffering from a bigger phase shift.

In our fault injector we have implemented a parametric multi-qubits fault functionality.
The user can select the number of qubits to be corrupted and the correspondent phase shifts.
Additionally, we have designed a procedure to identify the logical qubits of the circuit that are mapped in physically close qubits in the real machine.
Even if current prototypes of qubits have relative large sizes (as the first transistors), with the rapid shrink of qubits we expect multi-qubits faults to be even more common.


\section{Quantum Fault Injector}
\label{sec_fi}

In this Section we formalize the metric we use to understand if the injected fault affects the quantum circuit output, we present our quantum fault injector, QuFI, and describe its capabilities and potentialities. 

\subsection{Quantum Vulnerability Factor}

The output of a quantum circuit, as described in Section~\ref{sec_background}, is probabilistic. After the execution of a circuit, the result is one of the possible states. The correct output is(are) the state(s) with the higher probability. Thus, to measure the probability of each state, and select the correct state as output, one circuit needs to be executed several times. The standard number of times each circuit is executed for IBM/Qiskit is $1,024$, generating a probability distribution among the observed states. A state with 50\% probability will be observed about $512$ times. This also means that, to measure the impact of a fault, we need to run the injection several times (at least $1,024$) and check how the output probability distribution is changed.

To measure the impact of a fault on the output correctness (i.e., on the output probability distribution), we use the Quantum Vulnerability Factor (QVF) metric. As the AVF or PVF for classical computers~\cite{Mukherjee2003, PVF}, the QVF indicates the probability for the fault to propagate and affect the computation output.

\begin{figure*}[ht!]%
   	\centering
    	\includegraphics[width=0.98\textwidth]{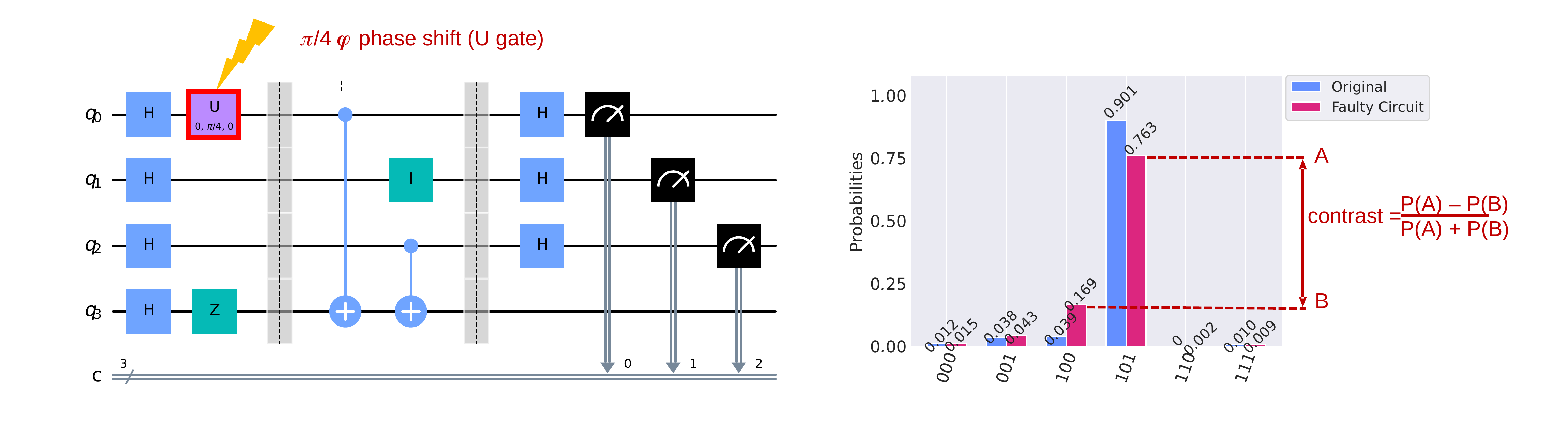}
        \caption{Example of fault injection in the Bernstein-Vazirani circuit (left) and QVF calculation (right). A $\theta$ shift of $\frac{\pi}{4}$ is injected in $q_0$ after the first H-gate. The fault modifies the output probabilities distribution, shown to the right, from the blue one to the red one. As shown in Equation~\ref{eq_qvf}, QVF is calculated using the Michelson Contrast, where A is the probability of the expected (fault-free) output (101 in this case) and B is the highest probability among the wrong outputs (100).}
        \label{fig_fault_example}
\end{figure*}

The QVF is calculated computing the Michelson Contrast~\cite{kukkonen1993michelson}, that measures how distinguishable one object is from others using color, luminance, or, as in our specific case, the probability of each output state.
In other words, as shown in Figure~\ref{fig_fault_example}, the Michelson Contrast defines how confidently (i.e., how distinguished) one can select the correct state among all states in the output. We recall that the quantum circuit output is a superposition of states, we need to pick the most probable one(s). The contrast is calculated with Equation~\ref{eq_contrast}:
\begin{equation}
    Contrast = \frac{P(A) - P(B)}{P(A) + P(B)}
    \label{eq_contrast}
\end{equation}
where $P(A)$ is the probability of the correct state (i.e., the expected state in a fault-free execution), and $P(B)$ is the highest probability among any of the incorrect states (i.e., the most probable incorrect state). 
The contrast metric is not limited to circuits with a single correct state, the extension for multiple-state circuits can be easily performed by aggregating the probabilities of all correct states into $P(A)$.

The contrast range is $[-1,1]$, since it is possible for a circuit to produce $P(A) < P(B)$.
To shift the range to $[0,1]$ and to have lower values indicating a more reliable configuration (as for AVF and PVF), the QVF is calculated as shown in Equation~\ref{eq_qvf}:

\begin{equation}
    QVF= 1 - (Contrast + 1) /2
        \label{eq_qvf}
\end{equation}

A QVF close to zero indicates a clear contrast between the correct state and the incorrect ones, with the correct state presenting the highest probability. Thus, the probability of the expected output state is very high compared to the other states. Values close to one indicate that the correct states are not even as probable as the incorrect ones, i.e., the incorrect state is likely to be selected. QVF values around $0.5$ mean that the correct state and at least one incorrect state have similar probabilities, which makes the identification of correct states dubious.

\subsection{Single Fault Injection}

Our fault injector is built on top of  the open-source and well-documented Qiskit framework~\cite{Qiskit}.
The fault injector operates over a Qiskit's \textit{QuantumCircuit} object to generate new circuits with the injected fault(s). As shown in Figure~\ref{fig_fault_example}, to modify the qubit(s) state(s) we apply a tuned stimulus using an additional gate (an \textit{injector}) to specific points of the quantum circuit. This injector gate can be seen as the lines of code necessary to modify the values of a register or the output of an instruction in most traditional fault injectors.

To model the injected fault that, as discussed in Section~\ref{sec_model}, can have different phase shifts of different magnitude, we use the \textit{U} gate as injector.
The \textit{U} gate, which is the most flexible one, can be described as: 
\begin{equation}
U(\theta, \phi, \lambda)=\left[\begin{array}{cc}\cos \left(\frac{\theta}{2}\right) & -e^{i \lambda} \sin \left(\frac{\theta}{2}\right) \\ e^{i \phi} \sin \left(\frac{\theta}{2}\right) & e^{i(\phi+\lambda)} \cos \left(\frac{\theta}{2}\right)\end{array}\right]
\end{equation}
which receives three parameters: 
\begin{itemize}
    \item $\phi$ is the angle defined in the $XY$ plane of the Bloch sphere (a rotation angle on the $Z$ axis);
    \item $\theta$ is the angle defined in the plane that includes the $Z$ axis and the vector representing the generic quantum state $\ket{\psi}$; 
    \item $\lambda$ is also a rotation on the $Z$ axis.
\end{itemize}

The parameters for the $U$ gate have been thus selected:
\begin{itemize}
    \item $\phi$ = $[0, 2\pi[$ each $15{^\circ}$;
    \item $\theta$ = $[0, \pi]$ each $15^{\circ}$;
    \item $\lambda$ = 0
\end{itemize}
This angles combination results in $312$ possible configurations (injections) for each position in the quantum circuit.

The new faulty circuits can be transpiled and executed just as one would execute a regular \textit{QuantumCircuit}. The framework is very flexible, and allows to execute the faulty circuits in physical IBM-Q machines as well as other simulators, or even export them as QASM \cite{cross2021openqasm} files to load and execute the circuits on different systems.

To have a full understanding of the effects of faults in quantum circuits execution, our fault injector can be executed in three different scenarios.
(1) Simulation without external noise, which is ideal but not realistic.
(2) Simulation of a physical machine, tuning the noise over which the fault is injected using the IBM-Q noise model. This represents a realistic environment based on actual quantum computers.
(3) Physical execution on the available IBM-Q machine.
In this paper, we only present data obtained with scenario (2) and (3) since scenario (1) cannot be achieved in the real world.

To have a broad evaluation of faults propagation, we inject faults after \textit{each} gate of the original circuit, simulating faults in each one of the circuit operations (see Figure~\ref{fig_fault_example}). Then, for each possible fault we execute the faulty circuit $1,024$ times to measure the probability distribution and QVF for that specific fault.

\begin{figure*}[th!]%
    \begin{subfigure}{.33\textwidth}
   	    \centering
        \includegraphics[width=\textwidth]{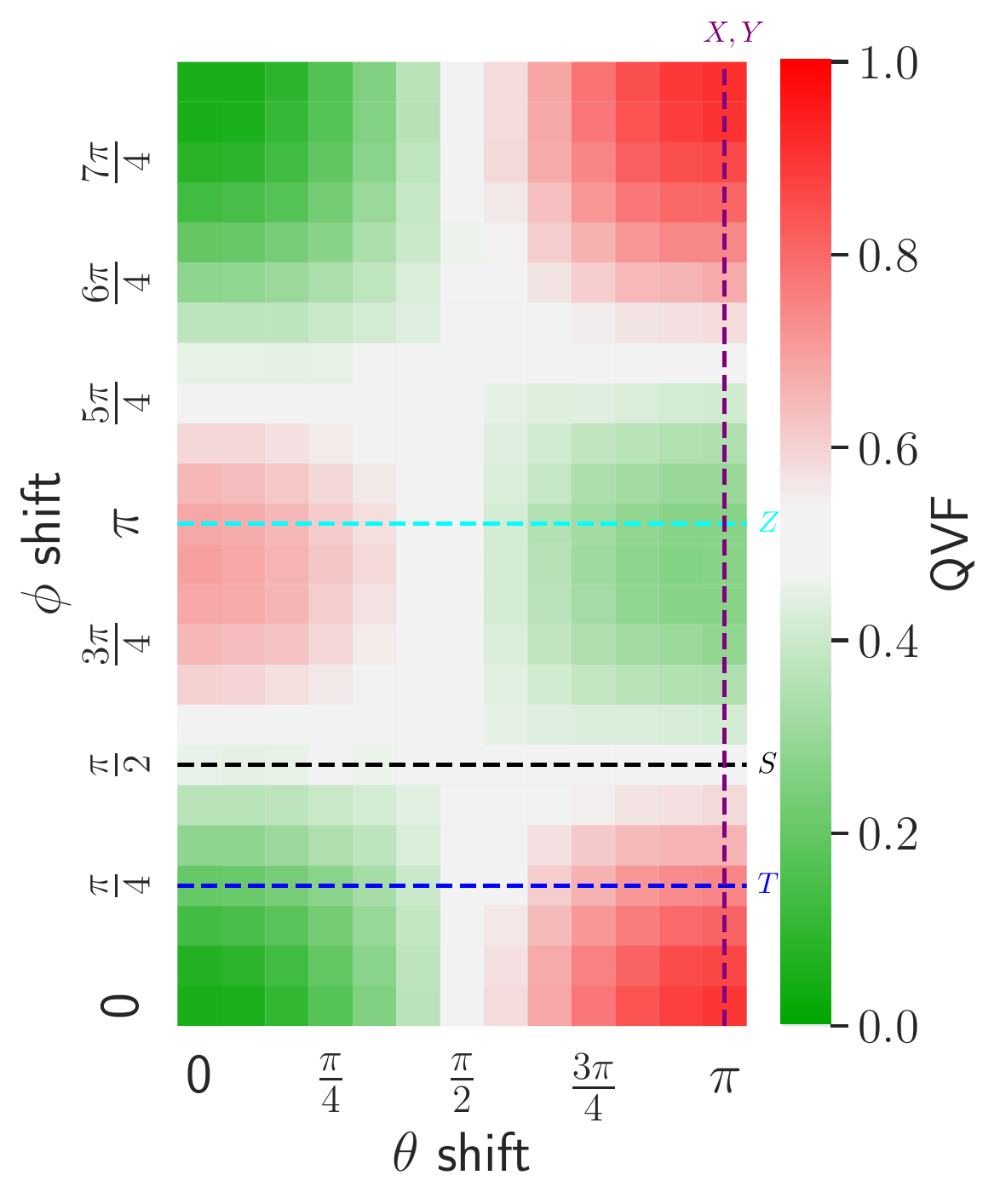}
        \caption{Bernstein-Vazirani.}
        \label{fig_hm_bv}
    \end{subfigure}%
    \hfill
    \begin{subfigure}{.33\textwidth}
        \centering
        \includegraphics[width=\textwidth]{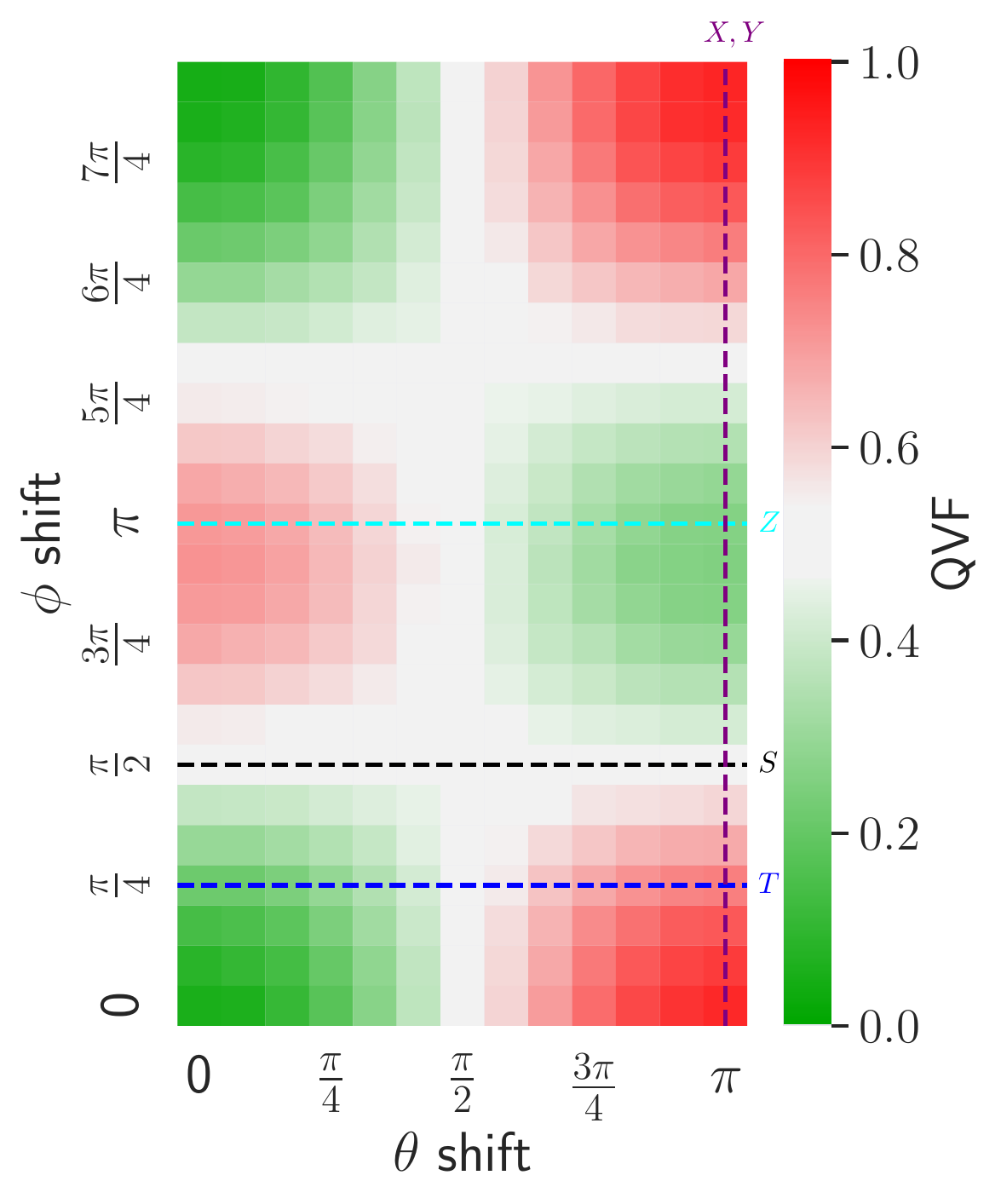}
        \caption{Deutsch-Jozsa.}
        \label{fig_hm_dj}
    \end{subfigure}%
    \hfill
    \begin{subfigure}{.33\textwidth}
        \centering
        \includegraphics[width=\textwidth]{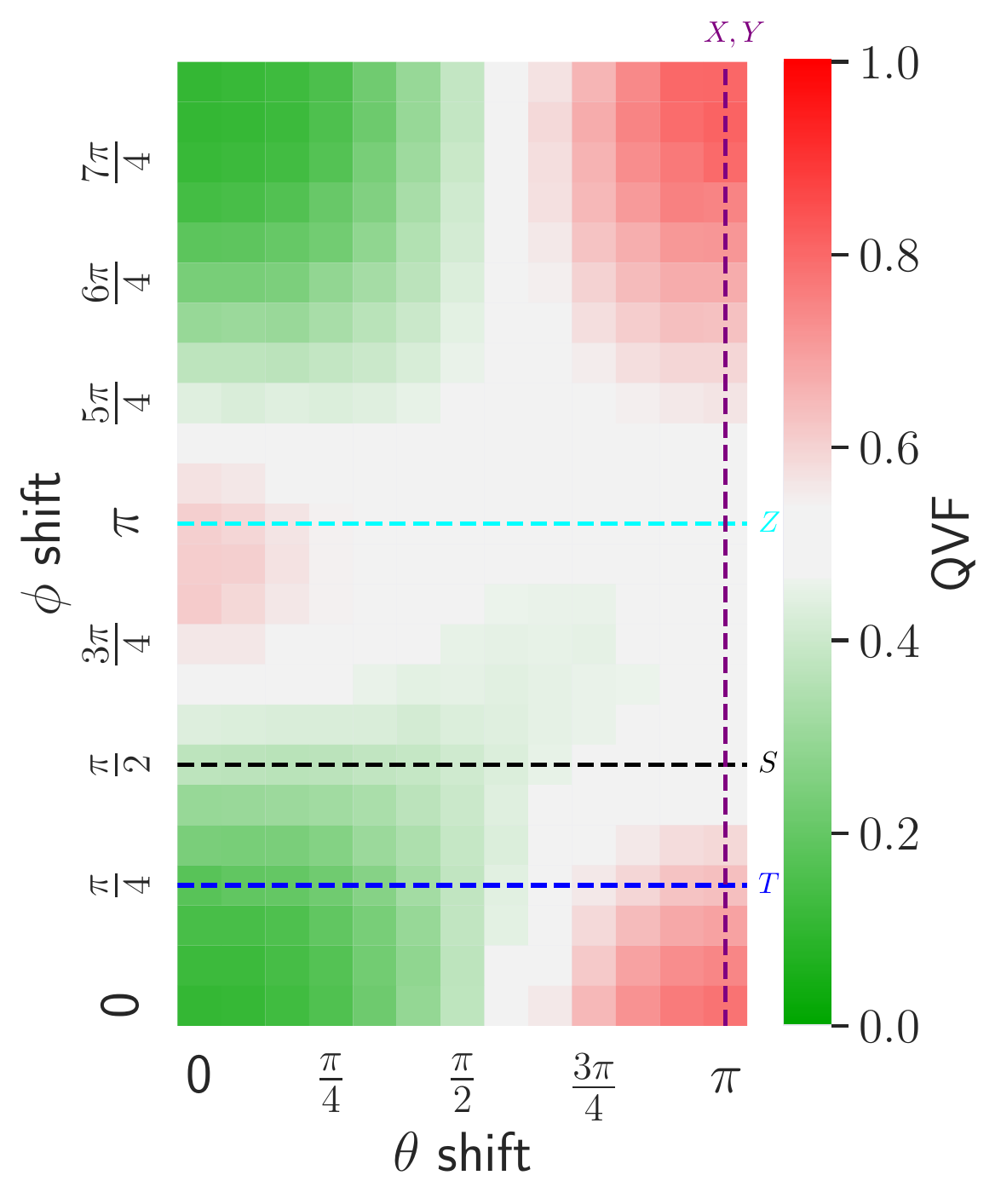}
        \caption{Quantum Fourier Transform.}
        \label{fig_hm_qft}
    \end{subfigure}%
    \caption{ \textbf{(a, b, c)} QVF heatmaps for the different 4-qubit circuits for different values of $\phi$ and/or $\theta$ shifts. The green color indicates a low QVF (the correct state can be confidently selected), the red color indicates a higher QVF (an incorrect output is more likely to be selected), and the white color indicates a dubious output (i.e., correct and incorrect states have about the same probability). The plot also shows dotted lines corresponding to the effect of common quantum gates ($X,Y,S,T,Z$) in order to provide a quick reference for the fault effect in the qubit.
    }
    \label{fig_heatmaps}
\end{figure*}

\subsection{Double Fault Injection}

It is known that a single particle strike can impact multiple qubits~\cite{muons2021}. As depicted in Figure~\ref{fig_ion}, the qubit closer to the particle impact will suffer a larger phase shift~\cite{Cardani2021}. Nonetheless, it is still unknown if the two superconducting qubits will be affected by a shift of the same phase and direction.

QuFI is highly flexible, and allows to inject any type of phase shift in any couple of qubits. For this paper, to study the impact of multi qubits faults, we decided to inject a first fault of a certain
magnitude 
on a certain qubit and a second fault of a smaller magnitude in a \textit{neighbor} qubit (a qubit physically close to the first one). In other words, the second qubit, which we suppose to be farther from the particle impact, will be affected by a phase shift, in any direction, that goes from 0 to the magnitude of the shift of the first fault.

In QuFI the first fault is injected as specified in the previous subsection, and, on top of that, we inject an additional fault on the neighboring qubits separately.
For ease of reading, we denote the first injection as ($\theta_{0}$,$\phi_{0}$) and the second injection as  ($\theta_{1}$,$\phi_{1}$).
The second injection is characterized by lower phase magnitude for the two angles, i.e. $\theta_{1}\le\theta_{0}$ and $\phi_{1}\le\phi_{0}$.

A key information for the double fault injection is the understanding where to perform ($\theta_{1}$,$\phi_{1}$) injection, i.e., to identify the qubits that are physically (not logically) close to each others. Intuitively, this is similar to memory in traditional computing, in which two bits that are logically next to each others are not necessarily physically close. 
The \textit{transpiling} (or \textit{transpilation} process) handles the logical-to-physical mapping of qubits, taking into account the machine topology.
We executed the transpiling process using the \textit{optimization\_level=3} in order to have the most dense layout and to reduce as much as possible the use of SWAP gates, which could change the ordering of qubits.
QuFI keeps track of the logical and physical qubits throughout the transpiling process, and tags the qubits that are neighbors after the transpiling process.


\section{Experimental Results}
\label{sec_results}

In this Section, we present the quantum circuits we tested and the three different evaluations we perform to demonstrate the flexibility of QuFI and to understand quantum circuits reliability. (1) We first evaluate circuits with a fixed size of 4 qubits, performing single fault injection. (2) Then, we consider the impact of the number of qubits in the circuit reliability, increasing the depth up to 7 qubits. (3) Finally, we inject double faults to assess the impact of correlated qubits corruptions.

\subsection{Tested Quantum Circuits}
\label{sub:circuits}

We consider three of the most used and widely known quantum circuits for our analysis.

\textbf{Deutsch-Josza} (DJ) circuit, given a function executed, is able to identify if the function is constant or balanced.
While of limited practical use, Deutsch-Josza was the first algorithm that showed that Quantum Computer could be faster than classical computers. 
\textbf{Bernstein-Vazirani} (BV) algorithm is an extension of Deutsch-Josza that identifies a string encoded in a function. 
\textbf{Quantum Fourier Transform} (QFT) is the quantum analogue of the discrete Fourier transform.
It applies a linear transformation to qubits and it is particularly interesting as it is a fundamental part of many quantum algorithms, such as Shor's factoring algorithm, Quantum Phase Estimation (QFE), and the computing of discrete logs.

\subsection{Fixed Width}

\begin{figure*}[th!]%
    \begin{subfigure}{.24\textwidth}
   	    \centering
        \includegraphics[width=\textwidth]{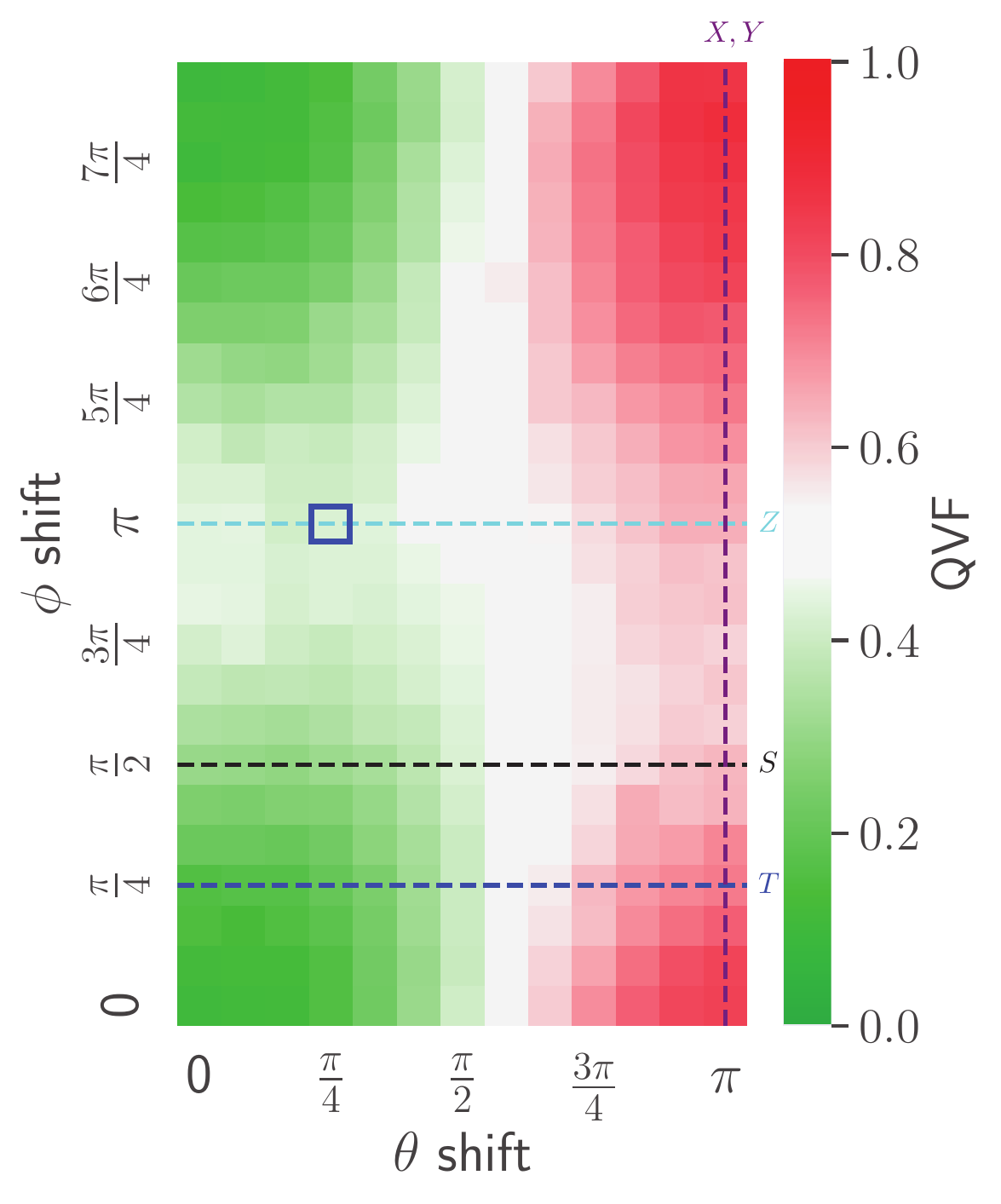}
        \caption{Qubit \#1}
        \label{fig_hm_qubits_1}
    \end{subfigure}%
    \hfill
    \begin{subfigure}{.24\textwidth}
        \centering
        \includegraphics[width=\textwidth]{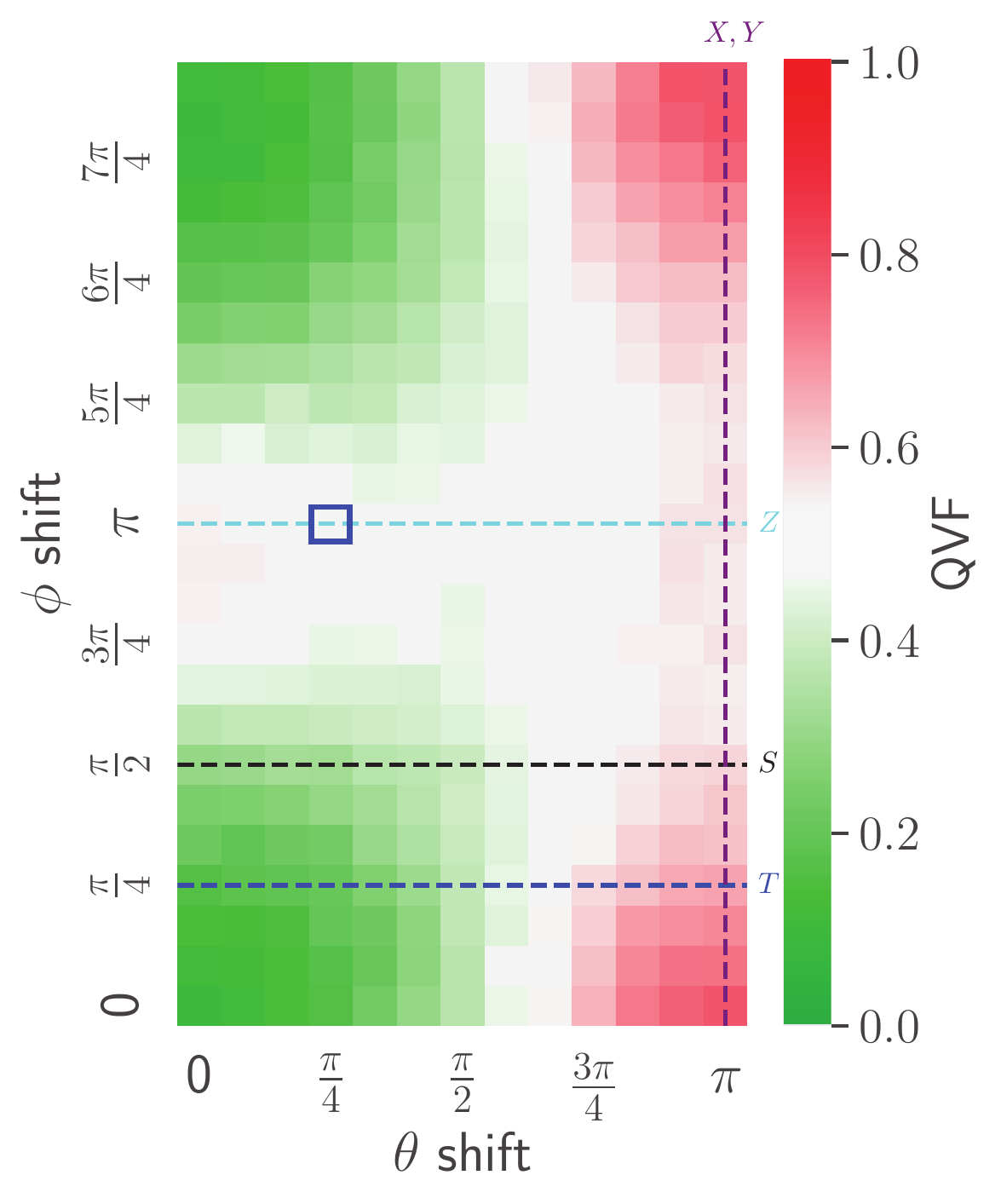}
        \caption{Qubit \#2}
        \label{fig_hm_qubits_2}
    \end{subfigure}%
    \hfill
    \begin{subfigure}{.24\textwidth}
        \centering
        \includegraphics[width=\textwidth]{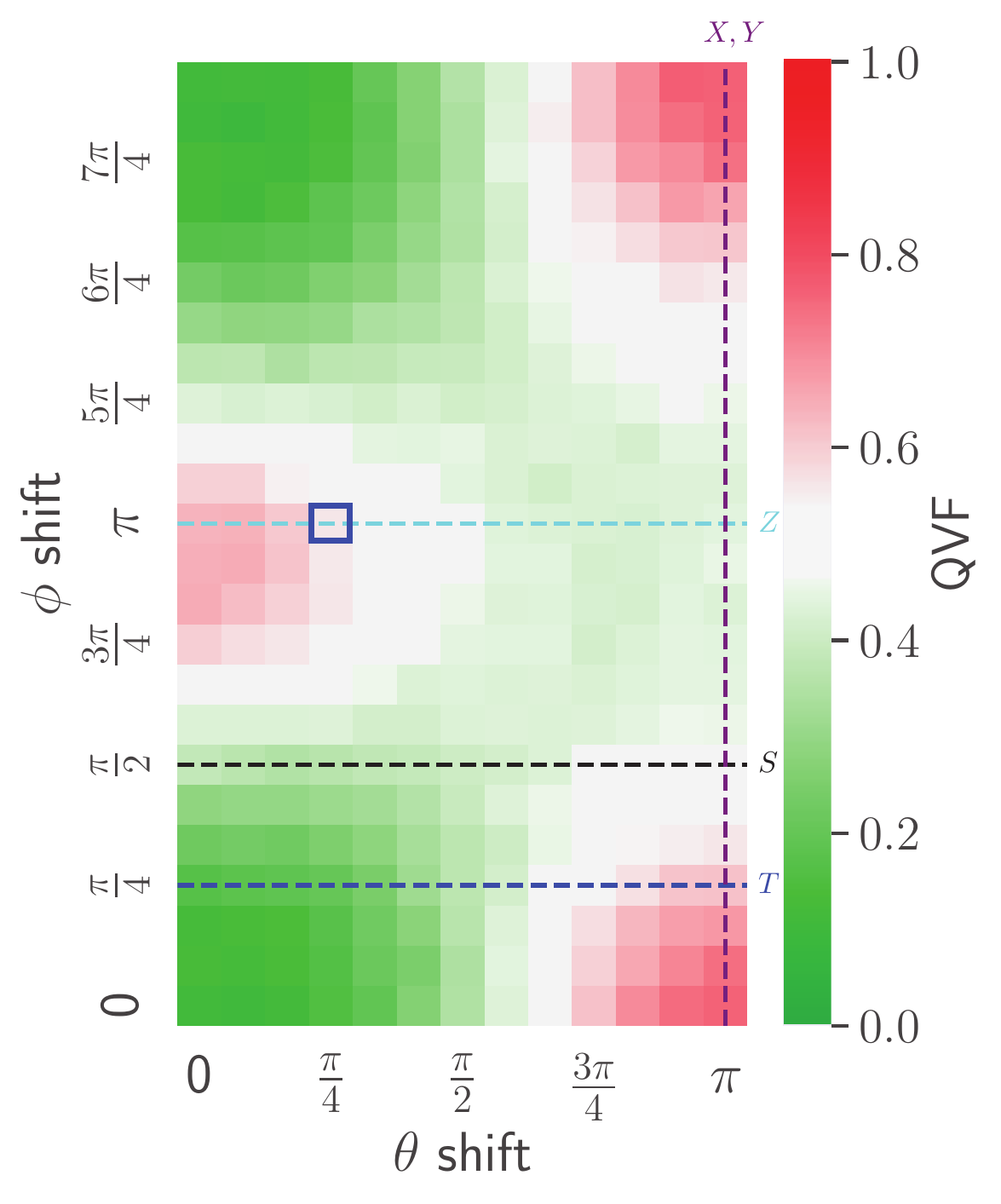}
        \caption{Qubit \#3}
        \label{fig_hmm_qubits_3}
    \end{subfigure}%
    \hfill
    \begin{subfigure}{.24\textwidth}
        \centering
        \includegraphics[width=\textwidth]{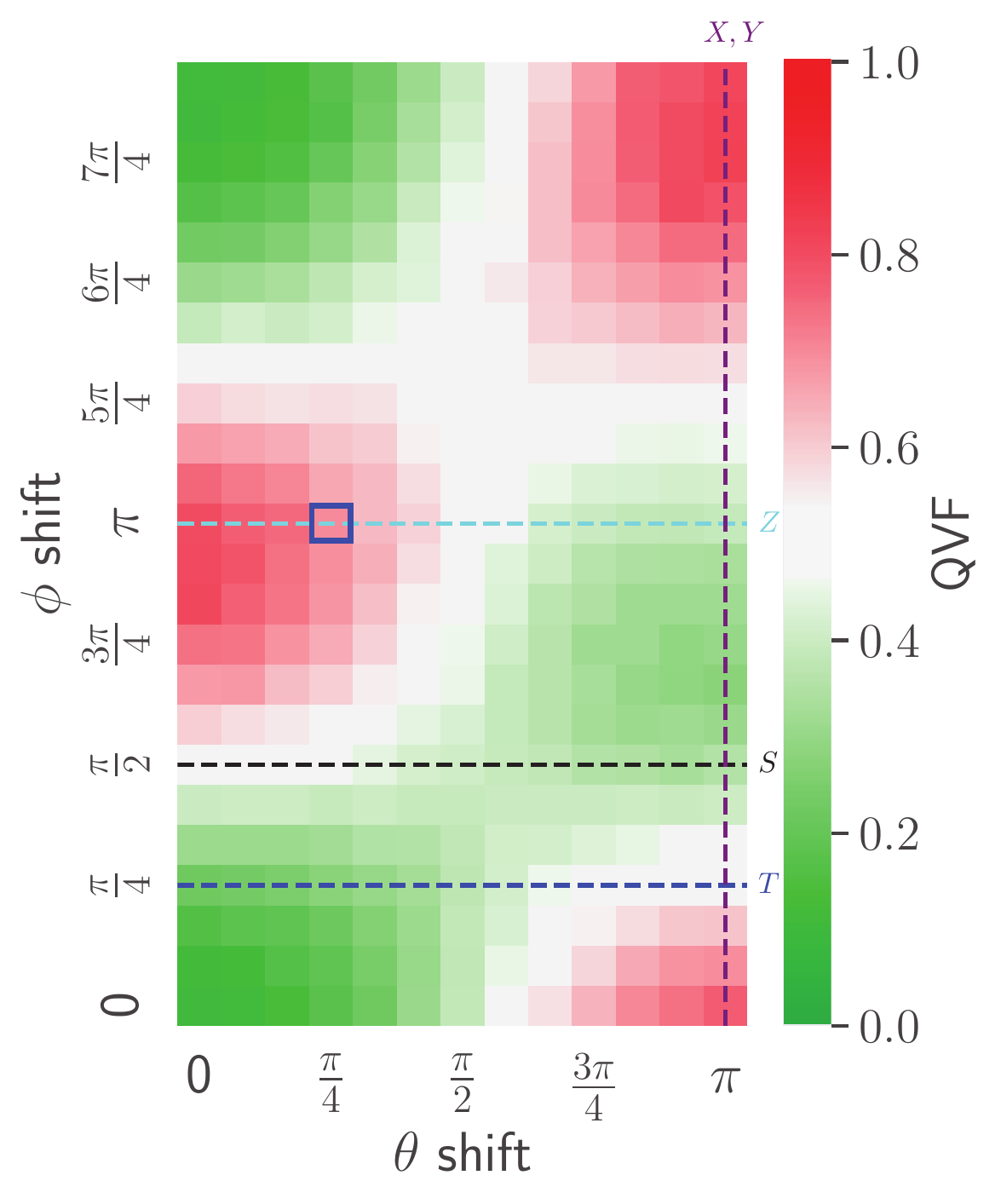}
        \caption{Qubit \#4}
        \label{fig_hmm_qubits_4}
    \end{subfigure}%
    \caption{ \textbf{(a, b, c, d)} QVF heatmaps of the single qubits for a 4-qubit implementation of the QFT algorithm.
    }
    \label{fig_heatmaps_qubits}
\end{figure*}

Since a qubit fault is not binary, as discussed in Section~\ref{sec_model}, we need to inject a great number of possible phase shifts for each fault location, and each injection can have various impacts on the circuit output. We present the results of a total of $18,849,792$ fault injections.
We start the fault injection results analysis by showing, in Figure~\ref{fig_heatmaps}, the heatmap of the QVF for the considered quantum circuits.
We plot, for each circuit, the QVF computed for each injected fault (i.e., the injected $\phi$ and/or $\theta$ shifts).
Each spot $(\phi,\theta)$ represents the QVF mean for all possible fault injections 
(qubit and position in the circuit depth) 
for that specific $(\phi,\theta)$ phase shift.
In Figure~\ref{fig_heatmaps}, we also superposed colored lines that correspond to the impact that would be imposed by common quantum gates ($X,Y,S,T,Z$) to the output value. E.g., a fault inducing a $\phi$ phase shift of $\pi$ is the equivalent of applying an additional $Z$ gate to the qubit. This is to better visualize the operative effect of the faults.

The colors we have selected in Figure~\ref{fig_heatmaps} help in quickly visualizing the effect of the injection.
Green (QVF $< 0.45$) indicates that the circuit correct output is still the most probable one, thus the fault has a minor impact. 
White ($0.45 <$ QVF $< 0.55$) indicates that the injection makes the output to be dubious (i.e., the correct output cannot be confidently selected).
Finally, red (QVF $> 0.55$) indicates that the injection effect is so high to make an incorrect output as the most probable one.
In other words, the green spots will be masked faults, harmless to the circuit execution.
Red spots are to be considered silent errors that need to be mitigated.  
Thus, circuits with a higher number of green and white spots should be more robust than circuits with a higher number of red spots.
White spots are interesting since the output lacks a distribution with a sufficiently high probability to be considered the correct one, i.e., there are, unexpectedly, two or more equally probable distributions.
These undefined output state can even be considered as \textit{detectable} errors, since a higher than expected number of high probability outputs is produced. 


It is worth noting that, to have a realistic evaluation, we are injecting faults over the intrinsic noise of current quantum computers.
This is why a fault-free execution can still produce incorrect outputs with a certain, albeit small, probability.
A fault-free execution in Figure~\ref{fig_heatmaps} is the spot in position {$(\phi=0, \theta=0)$}, and its color is not solid green (i.e., QVF $> 0$) due to noise.
Actually we found that, in some rare cases ($\sim$0.9\%), the injections improve the circuit QVF compared to the fault-free (but noisy) execution.
The injected fault basically compensates the noise effect, making the output state distribution clearer and closer to the ideal (noise and fault free) case. 


By varying only the $\ket{0}$-$\ket{1}$ probability on $\theta$ (i.e., the bottom line where {$(\phi=0, \theta=[0,\pi])$}), we will retain the original phase $\phi$ value to isolate the effect of a shift in $\theta$. 
The QVF quickly degrades in the vicinity of an orthogonal shift ($\frac{\pi}{2}$ or $90^\circ$) where the direction starts to flip.
As we move to a shift of {$(\phi=0, \theta=\pi)$}, we reach the worst QVF value by effectively reversing the $\ket{0}$-$\ket{1}$ probability.
We observe a similar behavior for the phase $\phi$, except that the QVF for {$(\phi=\pi, \theta=0)$} is not as high as on the $\ket{0}$-$\ket{1}$ probability shift of {$(\phi=0, \theta=\pi)$}.
Thus, a shift in $\theta$ (i.e., a shift in the $\ket{0}$-$\ket{1}$ state probability) is indeed more critical than a shift in $\phi$.

\begin{figure*}[!ht]%
    \begin{subfigure}{.3\textwidth}
    \captionsetup{justification=centering}
   	    \centering
        \includegraphics[width=\textwidth]{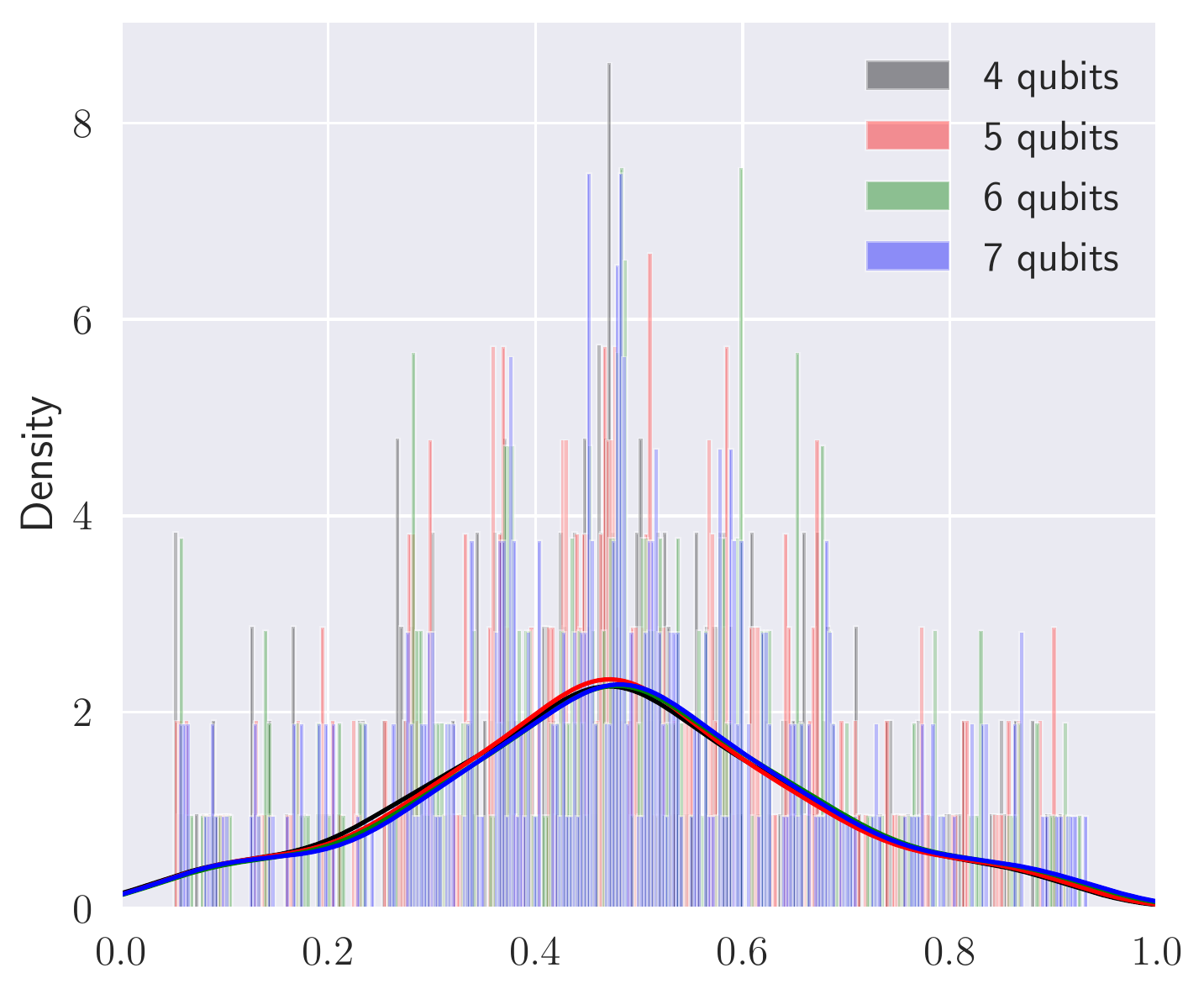}
        \caption{Bernstein-Vazirani}
        \label{fig_hist_depth_bv}
    \end{subfigure}%
    \hfill
    \begin{subfigure}{.3\textwidth}
    \captionsetup{justification=centering}
        \centering
        \includegraphics[width=\textwidth]{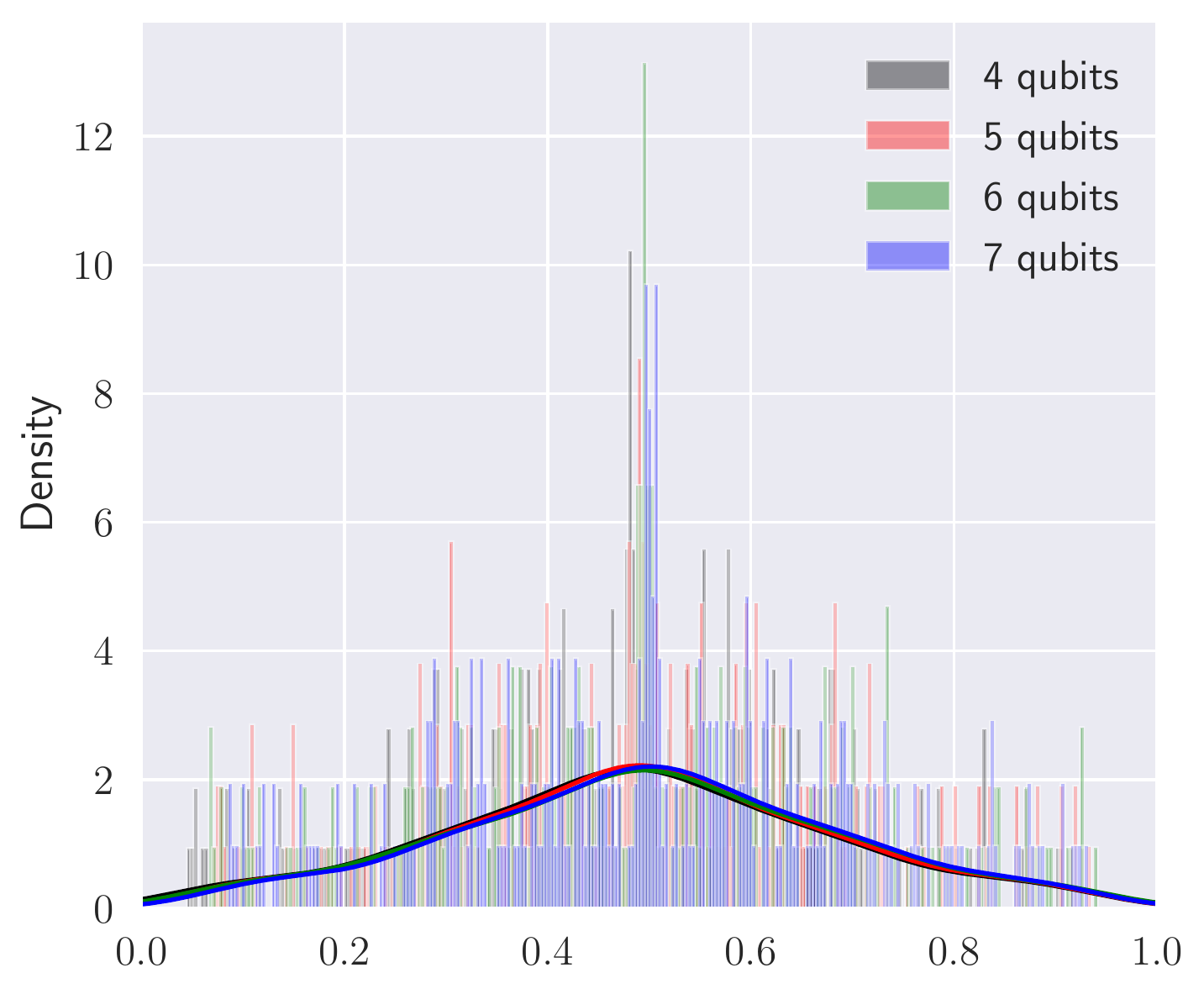}
        \caption{Deutsch-Jozsa}
        \label{fig_hist_depth_dj}
    \end{subfigure}%
    \hfill
    \begin{subfigure}{.3\textwidth}
    \captionsetup{justification=centering}
        \centering
        \includegraphics[width=\textwidth]{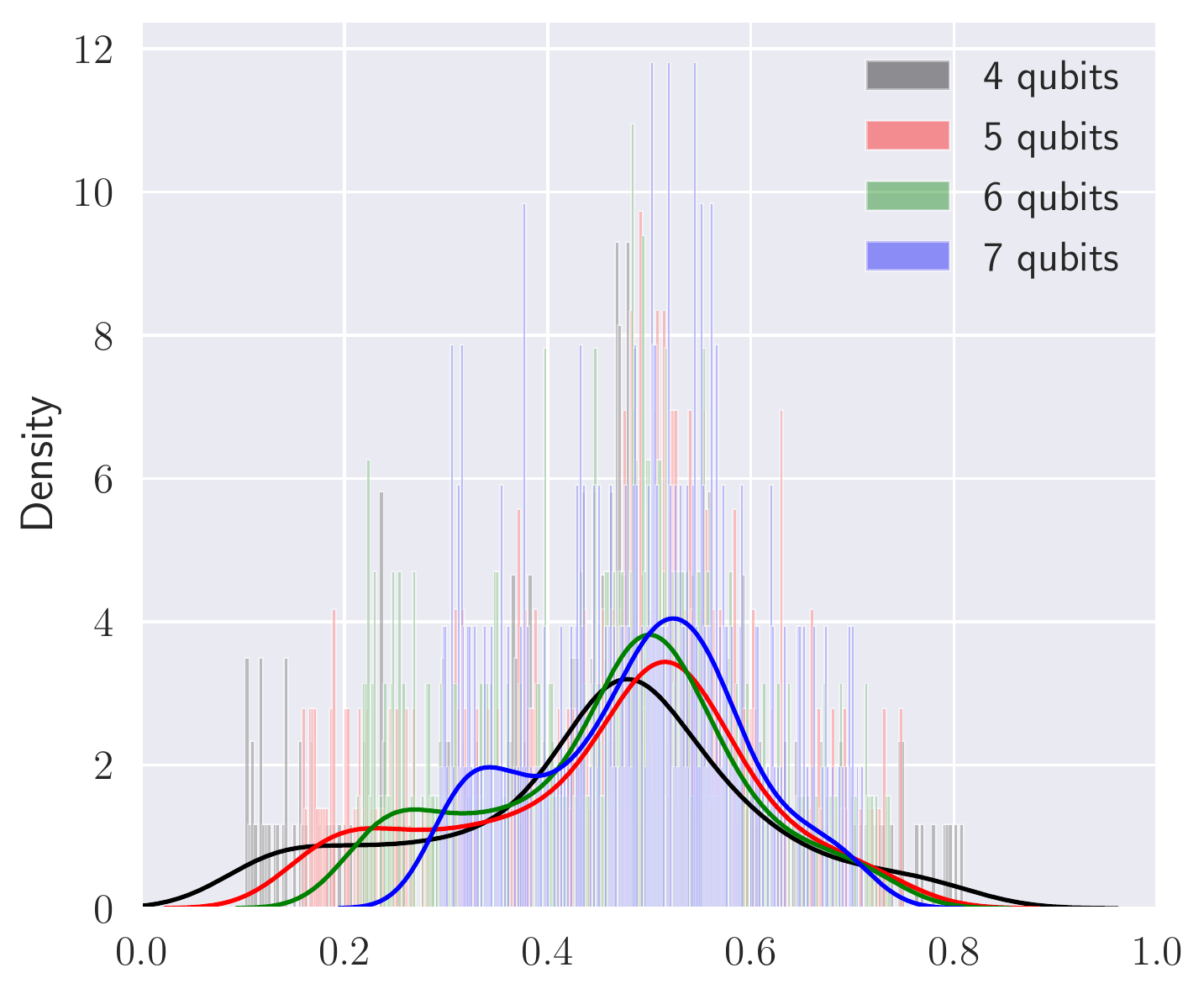}
        \caption{Quantum Fourier Transform}
        \label{fig_hist_depth_qft}
    \end{subfigure}%
    \caption{Histograms of the QVF distribution of the three considered circuits increasing circuit scale. For Bernstein-Vazirani \textbf{(a)} and Deutsch-Jozsa \textbf{(b)} the number of qubits does not modify the reliability profile. For QFT \textbf{(c)}, the higher the number of qubits the higher the number of harmful faults.
    }
    \label{fig_dist_circuit_depth}
\end{figure*}

It is worth noting that a combination of a $\theta$ and $\phi$ shift (e.g., $(\phi=\pi, \theta=\pi)$ does not produce unacceptable QVFs (red colors) for the circuits tested, as one could expect.
In fact, it has a tolerable effect for Bernstein-Vazirani and Deutsch-Jozsa resulting in acceptable QVFs (green colors).
This is not the case for QFT, though, demonstrating that the fault criticality is circuit-dependent.
For instance, a fault of $(\phi=\pi, \theta=\pi)$ is critical for QFT, but is harmless for Bernstein-Vazirani and Deutsch-Jozsa.

Interestingly, the QVF heatmaps in Figure~\ref{fig_heatmaps} show that the QVF, for Bernstein-Vazirani and Deutsch-Jozsa, is almost symmetric on $\phi$ with respect to $\pi$.
Actually, increasing the phase magnitude from $\pi$ towards $2\pi$ we are reversing the phase back to the original value of $\phi$.
This is not the case for Quantum Fourier Transform, that shows a not symmetric QVF, with faults on the diagonal from {$(\phi=0, \theta=0)$} to {$(\phi=\pi, \theta=\pi)$} having a lower QVF. As described next, and shown in Figure~\ref{fig_heatmaps_qubits}, each qubit presents a unique reliability profile, with some qubits showing a symmetric behavior while others qubits present a more complex behavior.

QuFI can also be used to assess the reliability of individual qubits. Similarly to AVF and PVF, one can use such information to design and implement extra fault tolerance solutions where they are more needed.
Additionally, some physical qubits may be more reliable than others due to the noise profile (included in our evaluation). Thus, the reliability information of individual logical qubits can also provide significant improvements for physical qubit mapping.
An example of this qubit assessment is represented by the heatmaps of Figure \ref{fig_heatmaps_qubits}, for the 4-qubit QFT circuit QVF. For lack of space we cannot report the individual QVF for the other circuits, we use QFT as a case study. Information about all the qubits of all the circuits can be found in our public repository~\cite{REPO}.
The profile of the QVF is different for the different qubits. To better convey this point, we take one phase shift as an illustrative example, which is the highlighted square representing the injection with a shift of $(\phi=\pi, \theta=\frac{\pi}{4})$.
The QVF value associated to this square is $0.4279$ for the first qubit, $0.4922$ for the second one, $0.5548$ for the third and $0.6909$ for the last qubit. Thus, the latter two qubits generates silent errors while the injection in the former one is masked, as it produces a sufficiently correct output.
This demonstrates how the fault effect can be significantly different on different qubits and proves how our framework can indeed be useful to assess their individual reliability.

Additionally, qubit 1 and 2 seem to tolerate faults as long as the $\theta$ shift is lower than $\frac{\pi}{2}$, and there is no tolerable effect as we increase the shift in both angles to $(\phi=\pi, \theta=\pi)$.
Qubit 3 and 4, in contrast, present even more tolerable effects by producing acceptable QVF even for $(\phi=\frac{\pi}{2}, \theta=\frac{\pi}{2})$.


\subsection{Circuit Scaling}



For quantum computers to compete with current supercomputers, solving real problems with the same scale (i.e., input size or precision) but faster than current classical performance (i.e., $10^{18}$ floating-point operations per second), they should have hundreds of qubits, depending on the problem class~\cite{Dalzell2020howmanyqubitsare}. 

For instance, Instantaneous Quantum Polynomial-Time (IQP) circuits requires 208 qubits while Quantum Approximate Optimization Algorithm (QAOA) requires 420 qubits. Moreover, Shor's algorithm, that can be used to crack RSA encryption, requires hundreds or even thousands of qubits to factor numbers using 256 up to 2048 bits~\cite{martin2012experimental,Gidney2021factorRSA, beauregard2002circuit}.
Thus, it is important to evaluate the circuits reliability behavior as we increase the circuit scale (i.e., the number of qubits and gates applied).


To understand if increasing the scale of a circuit impacts the fault propagation, we increase the number of qubits of the tested circuits from 4 up to 7, performing an additional fault injection campaign of $96,804,864$ faults.
We show the histogram of each circuit and scale in Figure~\ref{fig_dist_circuit_depth}. The black line depicts circuits using 4 qubits, red 5, green 6, and blue 7 qubits.
The QFT histogram, in contrast to Bernstein-Vazirani and Deutsch-Jozsa, has a skewed distribution toward the left side (lower QVF values), indicating a higher number of acceptable values (green spots) compared to unacceptable ones (red spots).
The three circuits have a mean value about $0.45$, which would be in the range of detectable errors.
However, QFT has a lower standard deviation (higher peak), indicating a higher number of detectable errors (white spots).
Histograms plotting and other image processing techniques are useful since they provide a method to outline the reliability of a circuit in a way that does not require human intervention.
Such image analysis methods could be applied to a large number of random circuits and/or specific faults.


\begin{figure*}[!ht]%
    \centering
        \includegraphics[width=1\textwidth]{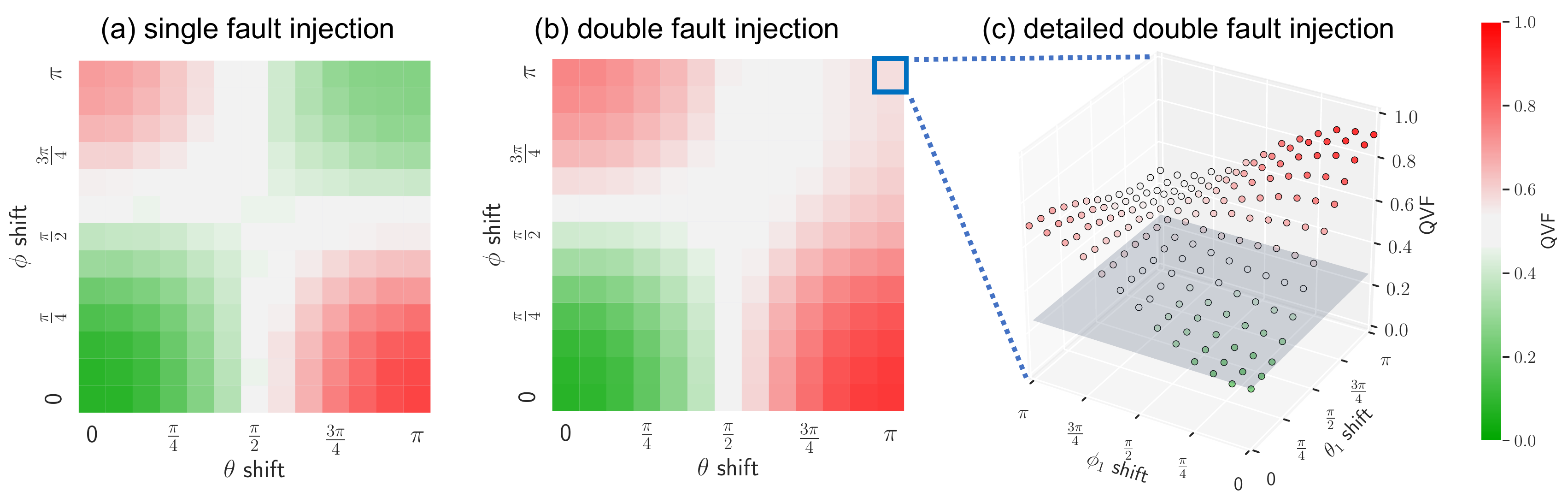}
    \caption{QVF for Bernstein-Vazirani for (a) Single and (b) Double fault injection. (c) Details of the QVFs for all the possible double faults injections, with the first fault injection fixed to $(\pi, \pi)$.}
    \label{fig_heatmaps_double}
\end{figure*}

\begin{figure}[!ht]%
    \centering
        \includegraphics[width=0.4\textwidth]{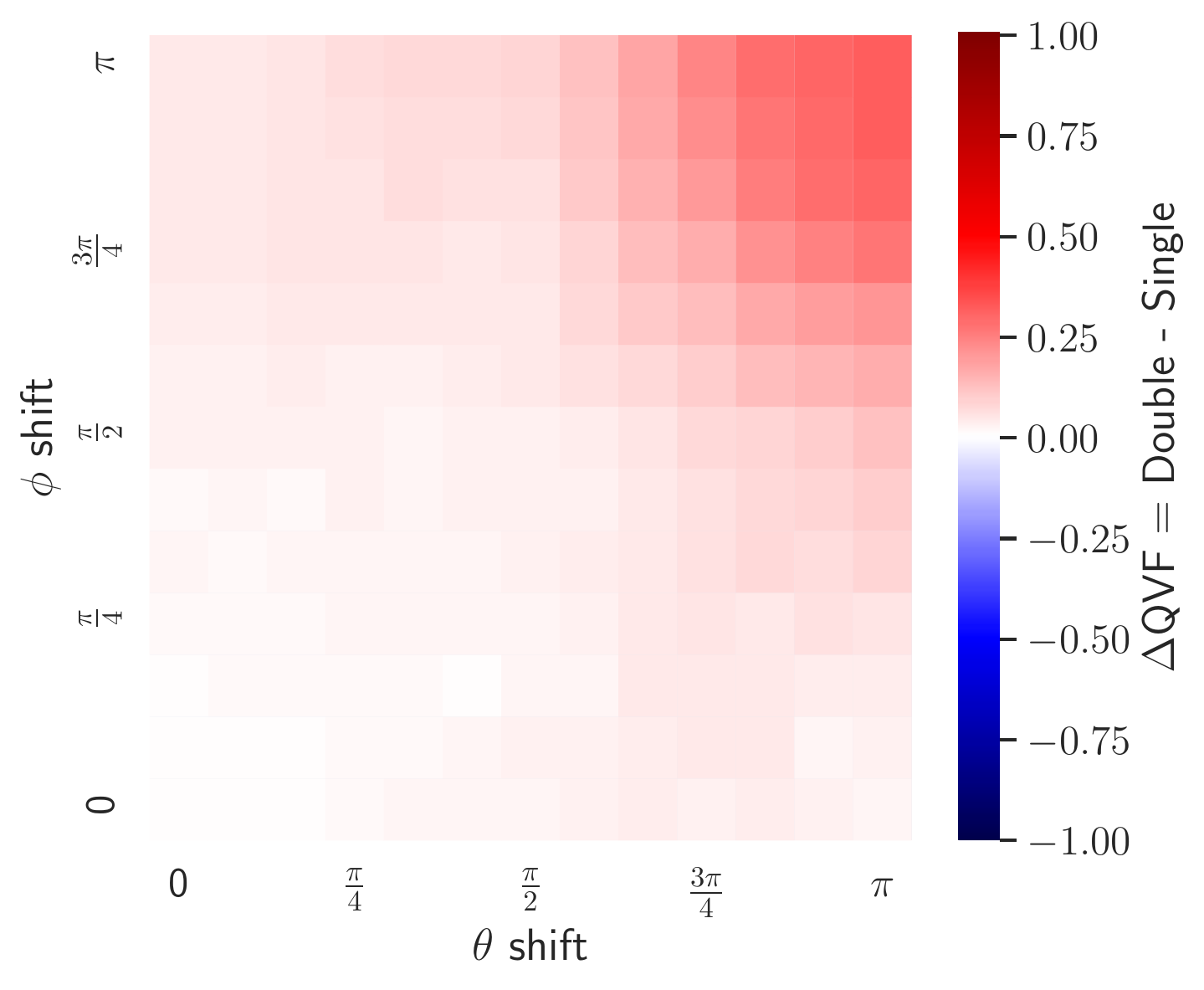}
    \caption{Comparison of the QVFs for Bernstein-Vazirani between single and double fault injection.}
    \label{fig_delta_single_double_bv}
\end{figure}

For Bernstein-Vazirani and Deutch-Jozsa, as shown in Figures~\ref{fig_hist_depth_bv} and~\ref{fig_hist_depth_dj}, the increase in circuit width and depth does not change the QVF, indicating that the scaling of these circuits does not impact the fault propagation. 
Obviously, increasing the circuit scale is likely to increase the error rate since more qubits are exposed. As any other work based on fault injection, in this paper we are evaluating the probability for a fault to propagate, assuming that a fault occurred. To measure the probability of fault occurrence in a circuit would require field or accelerate beam testing, which are out of the scope of this work. 

Interestingly, for QFT circuit, shown in Figure~\ref{fig_hist_depth_qft}, when we increase the number of qubit the QVF tends to the average value (i.e., lowering the standard deviation and increasing the peak around 0.5). Thus, as QFT circuit scales up, the number of harmless faults is reducing and the probability to have an output where the final user cannot confidently select the correct answer (e.g., $0.45 < QVF < 0.55$) increases.
It is worth noting that the scaling of QFT is particularly critical, as QFT is a fundamental piece of Shor's algorithm and many other algorithms.
Unless this effect is mitigated, the algorithm will hardly produce useful results when scaled up to hundreds of qubits.

For some circuits, then, the reliability profile can be scale-dependent. Thus, effective mitigation mechanisms for small scale circuits may be ineffective for large scaled ones, and the reliability profile should take into account the number of qubits as well as circuit depth.

\subsection{Multi Qubits Faults}

\begin{figure}[!ht]%
    \centering
        \includegraphics[width=0.45\textwidth]{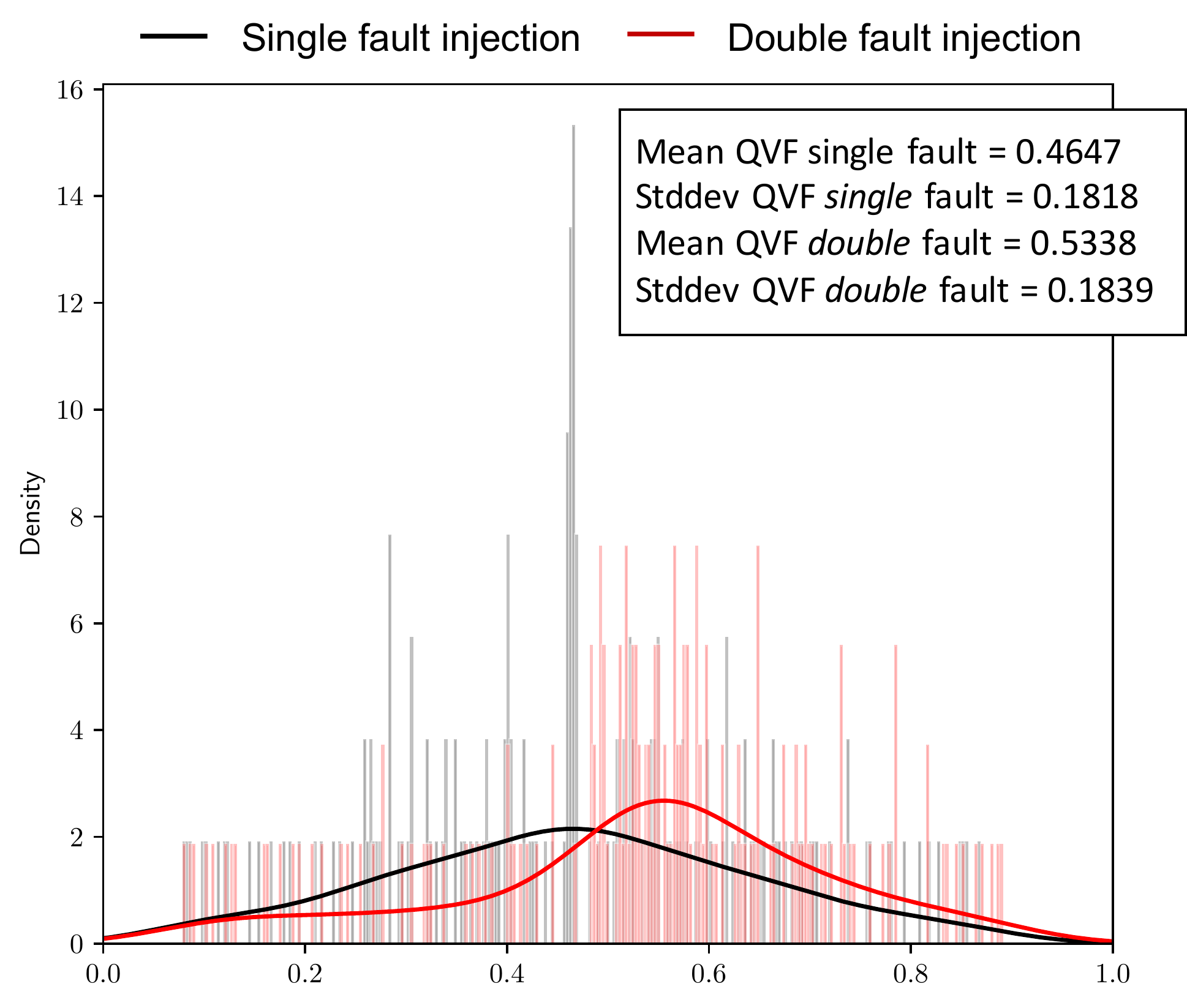}
    \caption{Bernstein-Vazirani distribution histogram considering all double faults combination.
    }
    \label{fig_dist_bv_double}
\end{figure}

In this subsection we describe the effect of injecting multiple faults in a circuit.
As described in Section~\ref{sub_multi}, we inject a fault in a qubit and an additional related fault of lower magnitude on the neighboring qubits. The set of all the possible qubit couples is identified during the \textit{transpilation} process, these couples will be the candidates on which the double injection tests will occur.

To illustrate the effect of this multi-injection, we use the Bernstein-Vazirani circuit as a case study, injecting a total of $169,594,880$ faults.
Figure \ref{fig_heatmaps_double} shows the effect of multi qubits injections on the circuit.
These plots show the heatmap of the QVF for Bernstein-Vazirani for (a) the single fault injection (it is portion of what is shown in Figure \ref{fig_heatmaps}, repeated here as a reference), (b) the double fault injection QVF (averaging the results obtained for all possible combinations of $\phi_{1}$ and $\theta_{1}$), and (c) a detailed view for all possible double fault injections with the first fault injection fixed to $(\phi_0 = \pi, \theta_0 = \pi)$.
For the sake of brevity we restrict the fault injection to $(0, \pi)$, since we have already seen (in Figure \ref{fig_heatmaps}) that for Bernstein-Vazirani the heatmaps are symmetric with respect to $\phi=\pi$.

Each square in Figure~\ref{fig_heatmaps_double}b for each $\theta_0$ and $\phi_0$ shift on the first qubit shows the \textit{average} among the QVFs of all the possible values of $\theta_1$ and $\phi_1$ shifts on the second qubit, with $\theta_1\le\theta_0$ and $\phi_1\le\phi_0$, as we assume the first qubit to be closer to the particle impact.
Figure \ref{fig_heatmaps_double}c shows, for the specific case of $\theta_0=\pi$ and $\phi_0=\pi$, the QVF for all the possible combinations of the second fault injections. 
All the explosion plots for the remaining cases are publicly available in \cite{REPO} for further analysis.




Not surprisingly, the second injection worsens (increases) the mean QVF (the number of red squares increases).
In particular, there is not the tolerable effect observed for the single fault injection in the case of $\theta_{0} = \pi$ and $\phi_{0} = \pi$ (i.e. there are no longer green squares on the top right corner of the figure). However, in Figure \ref{fig_heatmaps_double}a and \ref{fig_heatmaps_double}b it is still possible to observe a lower impact of the fault when both {$\theta$ and $\phi$} tend to $\pi$.
In general, the colors in the areas different than the {$\theta_{0}=0$,$\phi_{0}=0$} (top-left, top-right, bottom-right) show colors which tend to higher values of QVF, meaning that the circuit is more prone to have the (double) fault propagated to the output. 
This behavior is better highlighted in Figure \ref{fig_delta_single_double_bv}, which shows the $\Delta$QVF, i.e., the difference between the single and double fault injections QVF.
The QVF worsens, particularly when the phase shifts have higher magnitudes (close to $(\pi, \pi)$).

To have additional insights on the behavior of the double fault injection, Figure~\ref{fig_heatmaps_double}c shows the QVF for the Bernstein-Vazirani circuit obtained by fixing the phase shift in the first qubit to $\phi_{0}=\pi$ and $\theta_{0}=\pi$ and injecting in the second qubit a phase shift of $\phi_{1}\leq\pi$ and $\theta_{1}\leq\pi$.
In this sense it is a depiction with increased granularity of the highlighted square in Figure \ref{fig_heatmaps_double}b.
The gray plane indicates the QVF of the first fault without further injections and serves as a reference to show how much the QVF worsens when the second fault is injected.
It is possible to observe an interesting behavior which resembles that present in Figure \ref{fig_heatmaps_double}b (which gives a more general depiction of the faults effect on QVF) and that consists in a lower impact on QVF of the second injection when both $\phi_{1}$ and $\theta_{1}$ assume values closer to $\pi$, while the worst QVF values are obtained when only one of the two shifts is close to $\pi$ with the other tending to $0$.
In any case, for all possible configurations, the QVF remains worse than the single fault injection scenario.

Figure~\ref{fig_dist_bv_double} shows the distribution of QVF for the single (black color) and double (red color) fault injections on the Bernstein-Vazirani circuit.
The distribution related to the single fault injection has a mean QVF of $0.4647$ and a standard deviation of $0.1818$.
On the other hand, the distribution related to the double fault injection has a mean of $0.5338$ and a standard deviation of $0.1039$.
We can say that not only the red distribution has a higher mean, but also that it is more concentrated at higher values of QVF, supporting the claim that a double fault actually has a higher (negative) effect on the output.

\subsection{Physical Machine}

To demonstrate the QuFI versatility, we inject faults using a physical IBM quantum machine. We tested Bernstein-Vazirani on the IBM-Q Jakarta quantum machine. Due to time constraints for IBM physical machine reservations, we inject only four specific phase shift faults, which corresponds to basic gate operations (T, S, Z, and Y), in all possible fault positions, resulting in a total of $53,248$ faults injected. 

\begin{figure}[!t]%
    \centering
    \includegraphics[width=0.90\columnwidth]{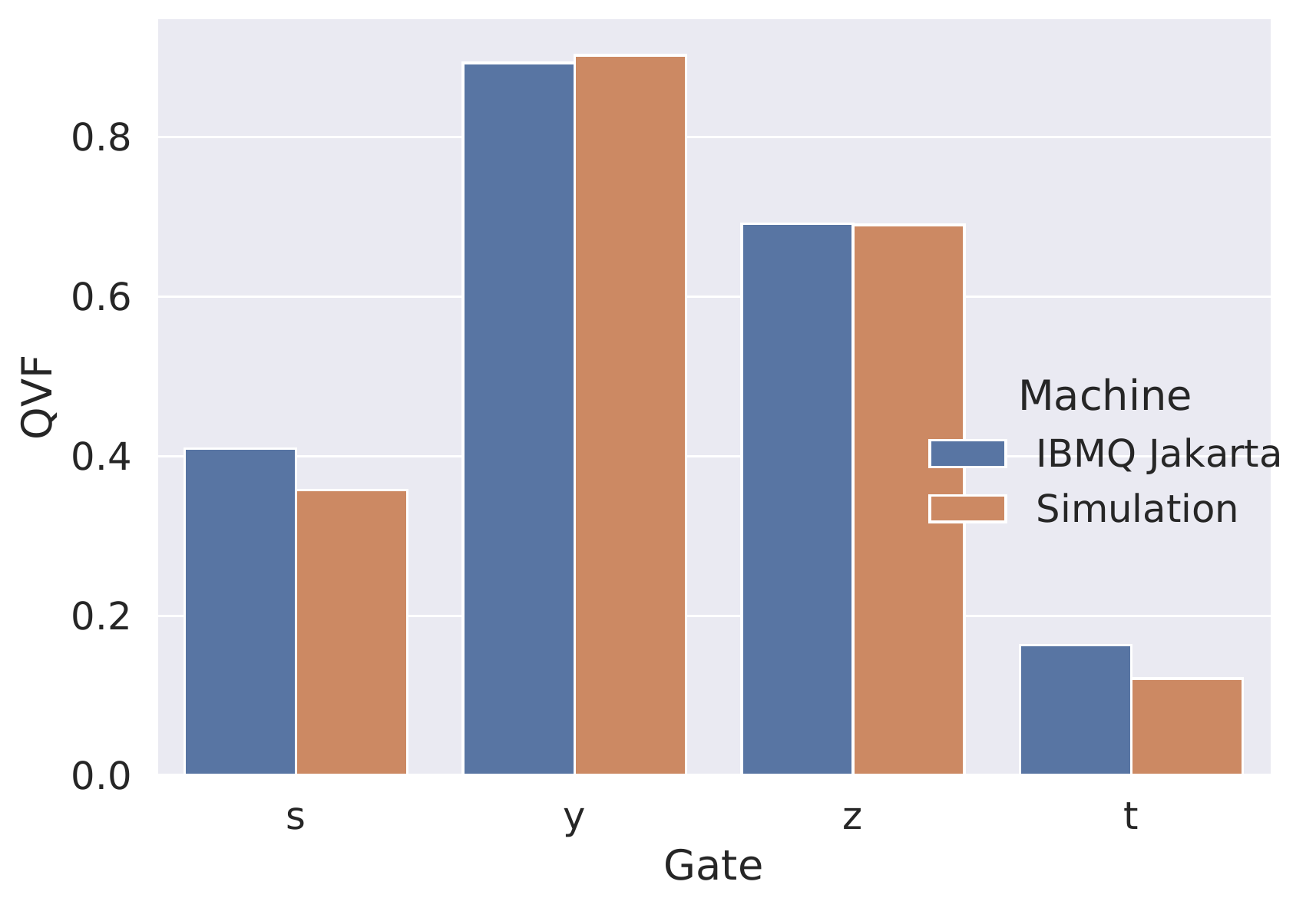}
        
    \caption{QVF comparison between simulation using IBM-Q noise model and physical machine execution (IBM-Q Jakarta).}
    \label{fig_histograms_real_simulation}
\end{figure}

We compare the results with a simulation including the IBM-Q noise model of the same machine. As we can see in Figure~\ref{fig_histograms_real_simulation}, there is only a small variation in QVF for the fault model types (e.g., absolute differences lower than $0.052$), which is expected since the noise is not static and may slightly change the state probability distribution. Thus, it is safe to assume that the results from simulation with noise models are precise enough to provide insights into physical machine executions.

\section{Conclusions}
\label{sec_conclusion}

In this paper, we have proposed the Quantum Fault Injector (QuFI) to better evaluate the sensitivity of quantum circuits and qubits to transient faults.
We have modeled transient faults as phase shifts based on the latest studies and radiation experiments performed on real quantum machines.
We consider both single and multiple faults which, according to~\cite{muons2021}, are going to be common in quantum computers.

We inject faults in three of the most used and widely known quantum circuits to demonstrate the flexibility of our fault injector. The evaluations and insights derived from the fault injection help identify the faults and qubits that are more likely to corrupt the output. Furthermore, we also evaluate the circuit scaling impact on reliability, demonstrating the need of our framework to help understand and mitigate the effects of transient faults.

\bibliographystyle{IEEEtran}
\bibliography{refs}

\begin{thebibliography}{10}
\providecommand{\url}[1]{#1}
\csname url@samestyle\endcsname
\providecommand{\newblock}{\relax}
\providecommand{\bibinfo}[2]{#2}
\providecommand{\BIBentrySTDinterwordspacing}{\spaceskip=0pt\relax}
\providecommand{\BIBentryALTinterwordstretchfactor}{4}
\providecommand{\BIBentryALTinterwordspacing}{\spaceskip=\fontdimen2\font plus
\BIBentryALTinterwordstretchfactor\fontdimen3\font minus
  \fontdimen4\font\relax}
\providecommand{\BIBforeignlanguage}[2]{{%
\expandafter\ifx\csname l@#1\endcsname\relax
\typeout{** WARNING: IEEEtran.bst: No hyphenation pattern has been}%
\typeout{** loaded for the language `#1'. Using the pattern for}%
\typeout{** the default language instead.}%
\else
\language=\csname l@#1\endcsname
\fi
#2}}
\providecommand{\BIBdecl}{\relax}
\BIBdecl

\bibitem{1996Grover}
L.~K. {Grover}, ``{A fast quantum mechanical algorithm for database search},''
  \emph{arXiv e-prints}, pp. quant--ph/9\,605\,043, May 1996.

\bibitem{lloyd2013quantum}
S.~Lloyd, M.~Mohseni, and P.~Rebentrost, ``Quantum algorithms for supervised
  and unsupervised machine learning,'' 2013.

\bibitem{Peruzzo2014}
\BIBentryALTinterwordspacing
A.~Peruzzo, J.~McClean, P.~Shadbolt, M.-H. Yung, X.-Q. Zhou, P.~J. Love,
  A.~Aspuru-Guzik, and J.~L. O'Brien, ``A variational eigenvalue solver on a
  photonic quantum processor,'' \emph{Nature Communications}, vol.~5, no.~1, p.
  4213, 2014. [Online]. Available: \url{https://doi.org/10.1038/ncomms5213}
\BIBentrySTDinterwordspacing

\bibitem{quantum-drug}
\BIBentryALTinterwordspacing
``Let's talk about quantum computing in drug discovery,'' \emph{C\&EN Global
  Enterprise}, vol.~98, no.~35, pp. 20--22, 09 2020. [Online]. Available:
  \url{https://doi.org/10.1021/cen-09835-feature2}
\BIBentrySTDinterwordspacing

\bibitem{Bravyi2018}
\BIBentryALTinterwordspacing
S.~Bravyi, M.~Englbrecht, R.~K{\"o}nig, and N.~Peard, ``Correcting coherent
  errors with surface codes,'' \emph{npj Quantum Information}, vol.~4, no.~1,
  p.~55, Oct 2018. [Online]. Available:
  \url{https://doi.org/10.1038/s41534-018-0106-y}
\BIBentrySTDinterwordspacing

\bibitem{Chamberland_2020}
\BIBentryALTinterwordspacing
C.~Chamberland, A.~Kubica, T.~J. Yoder, and G.~Zhu, ``Triangular color codes on
  trivalent graphs with flag qubits,'' \emph{New Journal of Physics}, vol.~22,
  no.~2, p. 023019, Feb 2020. [Online]. Available:
  \url{http://dx.doi.org/10.1088/1367-2630/ab68fd}
\BIBentrySTDinterwordspacing

\bibitem{Qiskit}
e.~a. MD~SAJID~ANIS, ``Qiskit: An open-source framework for quantum
  computing,'' 2021.

\bibitem{AngliQSimulator}
A.~Li, O.~Subasi, X.~Yang, and S.~Krishnamoorthy, \emph{Density Matrix Quantum
  Circuit Simulation via the BSP Machine on Modern GPU Clusters}.\hskip 1em
  plus 0.5em minus 0.4em\relax IEEE Press, 2020.

\bibitem{radiation2011}
\BIBentryALTinterwordspacing
A.~D. Corcoles, J.~M. Chow, J.~M. Gambetta, C.~Rigetti, J.~R. Rozen, G.~A.
  Keefe, M.~Beth~Rothwell, M.~B. Ketchen, and M.~Steffen, ``Protecting
  superconducting qubits from radiation,'' \emph{Applied Physics Letters},
  vol.~99, no.~18, p. 181906, 2011. [Online]. Available:
  \url{https://doi.org/10.1063/1.3658630}
\BIBentrySTDinterwordspacing

\bibitem{Cardani2021}
\BIBentryALTinterwordspacing
L.~Cardani, F.~Valenti, N.~Casali, G.~Catelani, T.~Charpentier, M.~Clemenza,
  I.~Colantoni, A.~Cruciani, G.~D'Imperio, L.~Gironi, L.~Gr{\"u}nhaupt,
  D.~Gusenkova, F.~Henriques, M.~Lagoin, M.~Martinez, G.~Pettinari, C.~Rusconi,
  O.~Sander, C.~Tomei, A.~V. Ustinov, M.~Weber, W.~Wernsdorfer, M.~Vignati,
  S.~Pirro, and I.~M. Pop, ``Reducing the impact of radioactivity on quantum
  circuits in a deep-underground facility,'' \emph{Nature Communications},
  vol.~12, no.~1, p. 2733, 2021. [Online]. Available:
  \url{https://doi.org/10.1038/s41467-021-23032-z}
\BIBentrySTDinterwordspacing

\bibitem{Martinis2021}
\BIBentryALTinterwordspacing
J.~M. Martinis, ``Saving superconducting quantum processors from decay and
  correlated errors generated by gamma and cosmic rays,'' \emph{npj Quantum
  Information}, vol.~7, no.~1, p.~90, 2021. [Online]. Available:
  \url{https://doi.org/10.1038/s41534-021-00431-0}
\BIBentrySTDinterwordspacing

\bibitem{Chen2021}
\BIBentryALTinterwordspacing
Z.~Chen, K.~J. Satzinger, J.~Atalaya, A.~N. Korotkov, A.~Dunsworth, D.~Sank,
  C.~Quintana, M.~McEwen, R.~Barends, P.~V. Klimov, S.~Hong, C.~Jones,
  A.~Petukhov, D.~Kafri, S.~Demura, B.~Burkett, C.~Gidney, A.~G. Fowler,
  A.~Paler, H.~Putterman, I.~Aleiner, F.~Arute, K.~Arya, R.~Babbush, J.~C.
  Bardin, A.~Bengtsson, A.~Bourassa, M.~Broughton, B.~B. Buckley, D.~A. Buell,
  N.~Bushnell, B.~Chiaro, R.~Collins, W.~Courtney, A.~R. Derk, D.~Eppens,
  C.~Erickson, E.~Farhi, B.~Foxen, M.~Giustina, A.~Greene, J.~A. Gross, M.~P.
  Harrigan, S.~D. Harrington, J.~Hilton, A.~Ho, T.~Huang, W.~J. Huggins, L.~B.
  Ioffe, S.~V. Isakov, E.~Jeffrey, Z.~Jiang, K.~Kechedzhi, S.~Kim, A.~Kitaev,
  F.~Kostritsa, D.~Landhuis, P.~Laptev, E.~Lucero, O.~Martin, J.~R. McClean,
  T.~McCourt, X.~Mi, K.~C. Miao, M.~Mohseni, S.~Montazeri, W.~Mruczkiewicz,
  J.~Mutus, O.~Naaman, M.~Neeley, C.~Neill, M.~Newman, M.~Y. Niu, T.~E.
  O'Brien, A.~Opremcak, E.~Ostby, B.~Pat{\'o}, N.~Redd, P.~Roushan, N.~C.
  Rubin, V.~Shvarts, D.~Strain, M.~Szalay, M.~D. Trevithick, B.~Villalonga,
  T.~White, Z.~J. Yao, P.~Yeh, J.~Yoo, A.~Zalcman, H.~Neven, S.~Boixo,
  V.~Smelyanskiy, Y.~Chen, A.~Megrant, J.~Kelly, and G.~Q. AI, ``Exponential
  suppression of bit or phase errors with cyclic error correction,''
  \emph{Nature}, vol. 595, no. 7867, pp. 383--387, 2021. [Online]. Available:
  \url{https://doi.org/10.1038/s41586-021-03588-y}
\BIBentrySTDinterwordspacing

\bibitem{LossMechanisms2018}
\BIBentryALTinterwordspacing
L.~Gr\"unhaupt, N.~Maleeva, S.~T. Skacel, M.~Calvo, F.~Levy-Bertrand, A.~V.
  Ustinov, H.~Rotzinger, A.~Monfardini, G.~Catelani, and I.~M. Pop, ``Loss
  mechanisms and quasiparticle dynamics in superconducting microwave resonators
  made of thin-film granular aluminum,'' \emph{Phys. Rev. Lett.}, vol. 121, p.
  117001, Sep 2018. [Online]. Available:
  \url{https://link.aps.org/doi/10.1103/PhysRevLett.121.117001}
\BIBentrySTDinterwordspacing

\bibitem{nature_rad}
\BIBentryALTinterwordspacing
A.~P. Veps{\"a}l{\"a}inen, A.~H. Karamlou, J.~L. Orrell, A.~S. Dogra, B.~Loer,
  F.~Vasconcelos, D.~K. Kim, A.~J. Melville, B.~M. Niedzielski, J.~L. Yoder,
  S.~Gustavsson, J.~A. Formaggio, B.~A. VanDevender, and W.~D. Oliver, ``Impact
  of ionizing radiation on superconducting qubit coherence,'' \emph{Nature},
  vol. 584, no. 7822, pp. 551--556, 2020. [Online]. Available:
  \url{https://doi.org/10.1038/s41586-020-2619-8}
\BIBentrySTDinterwordspacing

\bibitem{muons2021}
\BIBentryALTinterwordspacing
C.~D. Wilen, S.~Abdullah, N.~A. Kurinsky, C.~Stanford, L.~Cardani,
  G.~D'Imperio, C.~Tomei, L.~Faoro, L.~B. Ioffe, C.~H. Liu, A.~Opremcak, B.~G.
  Christensen, J.~L. DuBois, and R.~McDermott, ``Correlated charge noise and
  relaxation errors in superconducting qubits,'' \emph{Nature}, vol. 594, no.
  7863, pp. 369--373, 2021. [Online]. Available:
  \url{https://doi.org/10.1038/s41586-021-03557-5}
\BIBentrySTDinterwordspacing

\bibitem{Barends2011}
\BIBentryALTinterwordspacing
R.~Barends, J.~Wenner, M.~Lenander, Y.~Chen, R.~C. Bialczak, J.~Kelly,
  E.~Lucero, P.~O.~Malley, M.~Mariantoni, D.~Sank, H.~Wang, T.~C. White,
  Y.~Yin, J.~Zhao, A.~N. Cleland, J.~M. Martinis, and J.~J.~A. Baselmans,
  ``Minimizing quasiparticle generation from stray infrared light in
  superconducting quantum circuits,'' \emph{Applied Physics Letters}, vol.~99,
  no.~11, p. 113507, 2011. [Online]. Available:
  \url{https://doi.org/10.1063/1.3638063}
\BIBentrySTDinterwordspacing

\bibitem{Mukherjee2003}
S.~S. Mukherjee, C.~Weaver, J.~Emer, S.~K. Reinhardt, and T.~Austin, ``{A
  Systematic Methodology to Compute the Architectural Vulnerability Factors for
  a High-Performance Microprocessor},'' in \emph{Proceedings of the 36th Annual
  IEEE/ACM International Symposium on Microarchitecture}.\hskip 1em plus 0.5em
  minus 0.4em\relax Washington, DC, USA: IEEE Computer Society, 2003, pp. 29--.

\bibitem{PVF}
V.~{Sridharan} and D.~R. {Kaeli}, ``Eliminating microarchitectural dependency
  from architectural vulnerability,'' in \emph{2009 IEEE 15th International
  Symposium on High Performance Computer Architecture}, 2009, pp. 117--128.

\bibitem{trappedionlowdose}
\BIBentryALTinterwordspacing
J.~Cui, A.~J. Rasmusson, M.~Donofrio, Y.~Xie, E.~Wolanski, and P.~Richerme,
  ``Susceptibility of trapped-ion qubits to low-dose radiation sources,''
  \emph{Journal of Physics B: Atomic, Molecular and Optical Physics}, 2021.
  [Online]. Available:
  \url{http://iopscience.iop.org/article/10.1088/1361-6455/ac076c}
\BIBentrySTDinterwordspacing

\bibitem{qucloudHPCA}
L.~Liu and X.~Dou, ``Qucloud: A new qubit mapping mechanism for
  multi-programming quantum computing in cloud environment,'' in \emph{2021
  IEEE International Symposium on High-Performance Computer Architecture
  (HPCA)}, 2021, pp. 167--178.

\bibitem{Harper2020}
\BIBentryALTinterwordspacing
R.~Harper, S.~T. Flammia, and J.~J. Wallman, ``Efficient learning of quantum
  noise,'' \emph{Nature Physics}, vol.~16, no.~12, pp. 1184--1188, Dec 2020.
  [Online]. Available: \url{https://doi.org/10.1038/s41567-020-0992-8}
\BIBentrySTDinterwordspacing

\bibitem{2017APSMARR51007G}
D.~{Greenbaum} and Z.~{Dutton}, ``{Coherent errors in quantum error
  correction},'' in \emph{APS March Meeting Abstracts}, ser. APS Meeting
  Abstracts, vol. 2017, Mar. 2017, p. R51.007.

\bibitem{Hu01Decoherence}
X.~{Hu}, R.~{de Sousa}, and S.~D. {Sarma}, ``{Decoherence and Dephasing in
  Spin-Based Solid State Quantum Computers},'' in \emph{Foundations of Quantum
  Mechanics in the Light of New Technology ISQM-Tokyo '01}, Oct. 2002, pp.
  3--11.

\bibitem{Kjaergaard_2020}
\BIBentryALTinterwordspacing
M.~Kjaergaard, M.~E. Schwartz, J.~Braumüller, P.~Krantz, J.~I.-J. Wang,
  S.~Gustavsson, and W.~D. Oliver, ``Superconducting qubits: Current state of
  play,'' \emph{Annual Review of Condensed Matter Physics}, vol.~11, no.~1, p.
  369–395, Mar 2020. [Online]. Available:
  \url{http://dx.doi.org/10.1146/annurev-conmatphys-031119-050605}
\BIBentrySTDinterwordspacing

\bibitem{qce-2021}
\BIBentryALTinterwordspacing
Z.~Chen, K.~J. Satzinger, J.~Atalaya, A.~N. Korotkov, A.~Dunsworth, D.~Sank,
  C.~Quintana, M.~McEwen, R.~Barends, P.~V. Klimov, S.~Hong, C.~Jones,
  A.~Petukhov, D.~Kafri, S.~Demura, B.~Burkett, C.~Gidney, A.~G. Fowler,
  A.~Paler, H.~Putterman, I.~Aleiner, F.~Arute, K.~Arya, R.~Babbush, J.~C.
  Bardin, A.~Bengtsson, A.~Bourassa, M.~Broughton, B.~B. Buckley, D.~A. Buell,
  N.~Bushnell, B.~Chiaro, R.~Collins, W.~Courtney, A.~R. Derk, D.~Eppens,
  C.~Erickson, E.~Farhi, B.~Foxen, M.~Giustina, A.~Greene, J.~A. Gross, M.~P.
  Harrigan, S.~D. Harrington, J.~Hilton, A.~Ho, T.~Huang, W.~J. Huggins, L.~B.
  Ioffe, S.~V. Isakov, E.~Jeffrey, Z.~Jiang, K.~Kechedzhi, S.~Kim, A.~Kitaev,
  F.~Kostritsa, D.~Landhuis, P.~Laptev, E.~Lucero, O.~Martin, J.~R. McClean,
  T.~McCourt, X.~Mi, K.~C. Miao, M.~Mohseni, S.~Montazeri, W.~Mruczkiewicz,
  J.~Mutus, O.~Naaman, M.~Neeley, C.~Neill, M.~Newman, M.~Y. Niu, T.~E.
  O'Brien, A.~Opremcak, E.~Ostby, B.~Pat{\'o}, N.~Redd, P.~Roushan, N.~C.
  Rubin, V.~Shvarts, D.~Strain, M.~Szalay, M.~D. Trevithick, B.~Villalonga,
  T.~White, Z.~J. Yao, P.~Yeh, J.~Yoo, A.~Zalcman, H.~Neven, S.~Boixo,
  V.~Smelyanskiy, Y.~Chen, A.~Megrant, J.~Kelly, and G.~Q. AI, ``Exponential
  suppression of bit or phase errors with cyclic error correction,''
  \emph{Nature}, vol. 595, no. 7867, pp. 383--387, 2021. [Online]. Available:
  \url{https://doi.org/10.1038/s41586-021-03588-y}
\BIBentrySTDinterwordspacing

\bibitem{muons2011}
B.~D. Sierawski, R.~A. Reed, M.~H. Mendenhall, R.~A. Weller, R.~D. Schrimpf,
  S.-J. Wen, R.~Wong, N.~Tam, and R.~C. Baumann, ``Effects of scaling on
  muon-induced soft errors,'' in \emph{2011 International Reliability Physics
  Symposium}, 2011, pp. 3C.3.1--3C.3.6.

\bibitem{Resch2021}
S.~Resch, S.~Tannu, U.~R. Karpuzcu, and M.~Qureshi, ``A day in the life of a
  quantum error,'' \emph{IEEE Computer Architecture Letters}, vol.~20, no.~1,
  pp. 13--16, 2021.

\bibitem{Baumann2005}
R.~Baumann, ``Soft errors in advanced computer systems,'' \emph{IEEE Design
  Test of Computers}, vol.~22, no.~3, pp. 258--266, May 2005.

\bibitem{Catelani2011}
\BIBentryALTinterwordspacing
G.~Catelani, R.~J. Schoelkopf, M.~H. Devoret, and L.~I. Glazman, ``Relaxation
  and frequency shifts induced by quasiparticles in superconducting qubits,''
  \emph{Phys. Rev. B}, vol.~84, p. 064517, Aug 2011. [Online]. Available:
  \url{https://link.aps.org/doi/10.1103/PhysRevB.84.064517}
\BIBentrySTDinterwordspacing

\bibitem{Geant4}
J.~Allison, K.~Amako, J.~Apostolakis, H.~Araujo, P.~Arce~Dubois, M.~Asai,
  G.~Barrand, R.~Capra, S.~Chauvie, R.~Chytracek, G.~Cirrone, G.~Cooperman,
  G.~Cosmo, G.~Cuttone, G.~Daquino, M.~Donszelmann, M.~Dressel, G.~Folger,
  F.~Foppiano, J.~Generowicz, V.~Grichine, S.~Guatelli, P.~Gumplinger,
  A.~Heikkinen, I.~Hrivnacova, A.~Howard, S.~Incerti, V.~Ivanchenko,
  T.~Johnson, F.~Jones, T.~Koi, R.~Kokoulin, M.~Kossov, H.~Kurashige, V.~Lara,
  S.~Larsson, F.~Lei, O.~Link, F.~Longo, M.~Maire, A.~Mantero, B.~Mascialino,
  I.~McLaren, P.~Mendez~Lorenzo, K.~Minamimoto, K.~Murakami, P.~Nieminen,
  L.~Pandola, S.~Parlati, L.~Peralta, J.~Perl, A.~Pfeiffer, M.~Pia, A.~Ribon,
  P.~Rodrigues, G.~Russo, S.~Sadilov, G.~Santin, T.~Sasaki, D.~Smith,
  N.~Starkov, S.~Tanaka, E.~Tcherniaev, B.~Tome, A.~Trindade, P.~Truscott,
  L.~Urban, M.~Verderi, A.~Walkden, J.~Wellisch, D.~Williams, D.~Wright, and
  H.~Yoshida, ``Geant4 developments and applications,'' \emph{IEEE Transactions
  on Nuclear Science}, vol.~53, no.~1, pp. 270--278, 2006.

\bibitem{strike}
P.~Foulliat, ``Instruments for mev irradiation from ev to mev,''
  \url{http://sons.uniroma2.it/ericeneutronschool/wp-content/uploads/2017/11/Senesi_MeV_n_ERICE2018_v8.pdf},
  accessed: 2021-09-30.

\bibitem{Oldham2003}
T.~Oldham and F.~McLean, ``Total ionizing dose effects in mos oxides and
  devices,'' \emph{IEEE Transactions on Nuclear Science}, vol.~50, no.~3, pp.
  483--499, 2003.

\bibitem{kukkonen1993michelson}
H.~Kukkonen, J.~Rovamo, K.~Tiippana, and R.~N{\"a}s{\"a}nen, ``Michelson
  contrast, rms contrast and energy of various spatial stimuli at threshold,''
  \emph{Vision research}, vol.~33, no.~10, pp. 1431--1436, 1993.

\bibitem{cross2021openqasm}
A.~W. Cross, A.~Javadi-Abhari, T.~Alexander, N.~de~Beaudrap, L.~S. Bishop,
  S.~Heidel, C.~A. Ryan, J.~Smolin, J.~M. Gambetta, and B.~R. Johnson,
  ``Openqasm 3: A broader and deeper quantum assembly language,'' 2021.

\bibitem{REPO}
``Work repository,'' \url{https://github.com/link_to_repo_after_publication},
  December 2021.

\bibitem{Dalzell2020howmanyqubitsare}
\BIBentryALTinterwordspacing
A.~M. Dalzell, A.~W. Harrow, D.~E. Koh, and R.~L. La~Placa, ``How many qubits
  are needed for quantum computational supremacy?'' \emph{{Quantum}}, vol.~4,
  p. 264, May 2020. [Online]. Available:
  \url{https://doi.org/10.22331/q-2020-05-11-264}
\BIBentrySTDinterwordspacing

\bibitem{martin2012experimental}
E.~Martin-Lopez, A.~Laing, T.~Lawson, R.~Alvarez, X.-Q. Zhou, and J.~L.
  O'brien, ``Experimental realization of shor's quantum factoring algorithm
  using qubit recycling,'' \emph{Nature photonics}, vol.~6, no.~11, pp.
  773--776, 2012.

\bibitem{Gidney2021factorRSA}
\BIBentryALTinterwordspacing
C.~Gidney and M.~Ekerå, ``How to factor 2048 bit rsa integers in 8 hours using
  20 million noisy qubits,'' \emph{Quantum}, vol.~5, p. 433, Apr 2021.
  [Online]. Available: \url{http://dx.doi.org/10.22331/q-2021-04-15-433}
\BIBentrySTDinterwordspacing

\bibitem{beauregard2002circuit}
S.~Beauregard, ``Circuit for shor's algorithm using 2n+ 3 qubits,'' \emph{arXiv
  preprint quant-ph/0205095}, 2002.

\end{thebibliography}

\end{document}